\documentclass[onecollarge,numbook,smallextended]{svjour2}       

\smartqed  
\usepackage{graphicx}
 \usepackage{mathptmx}      
\usepackage{amssymb}
\usepackage{amsmath}
\usepackage{amsbsy}
\usepackage{bm}
\usepackage{color}
\usepackage{graphicx}
\usepackage[normalem]{ulem}
\usepackage{calrsfs}
\usepackage{cite}
\usepackage{hyperref}
\newcommand{\beq}{\begin{equation}}
\newcommand{\eeq}{\end{equation}}
\newcommand{\bq}{\begin{eqnarray}}
\newcommand{\eq}{\end{eqnarray}}
\newcommand{\bqn}{\begin{eqnarray*}}
\newcommand{\eqn}{\end{eqnarray*}}
\newcommand{\nn}{\nonumber\\}
\newcommand{\rr}{\mathbf{r}}
\newcommand{\rp}{R}
\newcommand{\rpp}{\bar{R}}
\newcommand{\rmd}{\mathrm{d}}
\newcommand{\rme}{\mathrm{e}}
\newcommand{\rmi}{\mathrm{i}}
\newcommand{\Or}{\mathcal{O}}
\newcommand{\onezerozero}{$1^{(00)}$}
\newcommand{\onezeroone}{$1^{(01)}$}
\newcommand{\oneoneone}{$1^{(11)}$}
\newcommand{\twozerozero}{$2^{(00)}$}
\newcommand{\twozeroone}{$2^{(01)}$}
\newcommand{\twooneone}{$2^{(11)}$}
\newcommand{\threezerozero}{$3^{(00)}$}
\newcommand{\threezeroone}{$3^{(01)}$}
\newcommand{\threeoneone}{$3^{(11)}$}
\newcommand{\red}[1]{{\color{red} #1}}

\usepackage{epic}

\newcommand{\xx}{5}
\newcommand{\yy}{-1.5}
\newcommand{\xxb}{35}

\newcommand{\xmaxfour}{85}
\newcommand{\xmaxthree}{65}
\newcommand{\xmaxtwo}{45}
\newcommand{\xmaxone}{25}
\newcommand{\ymax}{30}

\newcommand{\thickn}{1mm}
\newcommand{\thinn}{0.2mm}

\newcommand{\circulo}[1]{{\put(#1,0){\circle*{9}}\put(#1,0){\circle{18}}}}

\newcommand{\liner}{\red{\line(1,0){42}}}

\newcommand{\linerr}{\red{\line(1,0){10}}}

\newcommand{\pthreeexact}{\begin{picture}(\xmaxthree,\ymax)(-\xxb,\yy)
\setlength{\unitlength}{.1mm}
\circulo{0}
\put(-55,15){$N-1$}
\circulo{60}
\put(45,15){$N$}
\put(120,0){\circle{18}}
\put(110,15){1}
\circulo{180}
\put(170,15){2}
\circulo{240}
\put(230,15){3}
\put(300,0){\circle{18}}
\put(290,15){4}
\linethickness{\thinn}
\qbezier(-112,8)(-60,90)(-8,8)
\qbezier(-52,-8)(0,-90)(52,-8)
\qbezier(8,8)(60,90)(112,8)
\qbezier(128,8)(180,90)(232,8)
\qbezier(248,8)(300,90)(352,8)
\qbezier(68,-8)(120,-90)(172,-8)
\qbezier(188,-8)(240,-90)(292,-8)
\qbezier(308,-8)(360,-90)(412,-8)
\multiput(-9, 0)(-16,0){7}{\line(-1,0){10}}
\multiput(308, 0)(16,0){7}{\linerr}
\linethickness{\thickn}
\put(9,0){\liner}
\put(69,0){\liner}
\put(129, 0){\liner}
\put(189,0){\liner}
\put(249,0){\liner}
\end{picture}}

\newcommand{\ponezerozero}{\begin{picture}(\xmaxone,\ymax)(-\xx,\yy)
\setlength{\unitlength}{.1mm}
\put(0,0){\circle{18}}
\put(-10,15){1}
\put(60,0){\circle{18}}
\put(50,15){2}
\linethickness{\thickn}
\multiput(9, 0)(16,0){3}{\linerr}
\end{picture}}

\newcommand{\ponezerozerobis}{\begin{picture}(\xmaxone,\ymax)(-\xx,\yy)
\setlength{\unitlength}{.1mm}
\put(0,0){\circle{18}}
\put(-10,15){2}
\put(60,0){\circle{18}}
\put(50,15){3}
\linethickness{\thickn}
\multiput(9, 0)(16,0){3}{\linerr}
\end{picture}}

\newcommand{\ponezeroone}{\begin{picture}(\xmaxtwo,\ymax)(-\xx,\yy)
\setlength{\unitlength}{.1mm}
\put(0,0){\circle{18}}
\put(-10,15){1}
\put(60,0){\circle{18}}
\put(50,15){2}
\circulo{120}
\put(110,15){3}
\linethickness{\thinn}
\qbezier(8,8)(60,90)(112,8)
\linethickness{\thickn}
\multiput(9, 0)(16,0){3}{\linerr}
\put(69,0){\liner}
\end{picture}}

\newcommand{\monezeroone}{\begin{picture}(\xmaxtwo,\ymax)(-\xx,\yy)
\setlength{\unitlength}{.1mm}
\put(0,0){\circle{18}}
\put(-10,15){1}
\put(60,0){\circle{18}}
\put(50,15){2}
\circulo{120}
\put(110,15){3}
\linethickness{\thinn}
\qbezier(8,8)(60,90)(112,8)
\linethickness{\thickn}
\put(69,0){\liner}
\end{picture}}

\newcommand{\ponezeroonebis}{\begin{picture}(\xmaxtwo,\ymax)(-\xx,\yy)
\setlength{\unitlength}{.1mm}
\put(0,0){\circle{18}}
\put(-10,15){2}
\put(60,0){\circle{18}}
\put(50,15){3}
\circulo{120}
\put(110,15){4}
\linethickness{\thinn}
\qbezier(8,-8)(60,-90)(112,-8)
\linethickness{\thickn}
\multiput(9, 0)(16,0){3}{\linerr}
\put(69,0){\liner}
\end{picture}}

\newcommand{\ponezeroonebisbis}{\begin{picture}(\xmaxtwo,\ymax)(-\xx,\yy)
\setlength{\unitlength}{.1mm}
\circulo{0}
\put(-10,15){$N$}
\put(60,0){\circle{18}}
\put(50,15){1}
\put(120,0){\circle{18}}
\put(110,15){2}
\linethickness{\thinn}
\qbezier(8,-8)(60,-90)(112,-8)
\linethickness{\thickn}
\multiput(69, 0)(16,0){3}{\linerr}
\put(9,0){\liner}
\end{picture}}

\newcommand{\poneoneone}{\begin{picture}(\xmaxthree,\ymax)(-\xx,\yy)
\setlength{\unitlength}{.1mm}
\circulo{0}
\put(-15,15){$N$}
\put(60,0){\circle{18}}
\put(50,15){1}
\put(120,0){\circle{18}}
\put(110,15){2}
\circulo{180}
\put(170,15){3}
\linethickness{\thinn}
\qbezier(8,-8)(60,-90)(112,-8)
\qbezier(68,8)(120,90)(172,8)
\linethickness{\thickn}
\put(9,0){\liner}
\put(129,0){\liner}
\multiput(69, 0)(16,0){3}{\linerr}
\end{picture}}

\newcommand{\ptwozerozero}{\begin{picture}(\xmaxtwo,\ymax)(-\xx,\yy)
\setlength{\unitlength}{.1mm}
\put(0,0){\circle{18}}
\put(-10,15){1}
\circulo{60}
\put(50,15){2}
\put(120,0){\circle{18}}
\put(110,15){3}
\linethickness{\thinn}
\qbezier[20](8,8)(60,90)(112,8)
\linethickness{\thickn}
\put(69,0){\liner}
\put(9,0){\liner}
\end{picture}}

\newcommand{\eonethree}{\begin{picture}(\xmaxtwo,\ymax)(-\xx,\yy)
\setlength{\unitlength}{.1mm}
\put(0,0){\circle{18}}
\put(-10,15){1}
\put(120,0){\circle{18}}
\put(110,15){3}
\linethickness{\thinn}
\qbezier[20](8,8)(60,90)(112,8)
\end{picture}}

\newcommand{\ptwozeroone}{\begin{picture}(\xmaxthree,\ymax)(-\xx,\yy)
\setlength{\unitlength}{.1mm}
\put(0,0){\circle{18}}
\put(-15,15){1}
\circulo{60}
\put(50,15){2}
\put(120,0){\circle{18}}
\put(110,15){3}
\circulo{180}
\put(170,15){4}
\linethickness{\thinn}
\qbezier[20](8,8)(60,90)(112,8)
\qbezier(68,-8)(120,-90)(172,-8)
\linethickness{\thickn}
\put(9,0){\liner}
\put(129,0){\liner}
\put(69,0){\liner}
\end{picture}}

\newcommand{\ptwooneone}{\begin{picture}(\xmaxfour,\ymax)(-\xx,\yy)
\setlength{\unitlength}{.1mm}
\circulo{0}
\put(-15,15){$N$}
\put(60,0){\circle{18}}
\put(50,15){1}
\circulo{120}
\put(110,15){2}
\put(180,0){\circle{18}}
\put(170,15){3}
\circulo{240}
\put(230,15){4}
\linethickness{\thinn}
\qbezier(8,-8)(60,-90)(112,-8)
\qbezier[20](68,8)(120,90)(172,8)
\qbezier(128,-8)(180,-90)(232,-8)
\linethickness{\thickn}
\put(9,0){\liner}
\put(129,0){\liner}
\put(69,0){\liner}
\put(189,0){\liner}
\end{picture}}

\newcommand{\pthreezerozero}{\begin{picture}(\xmaxthree,\ymax)(-\xx,\yy)
\setlength{\unitlength}{.1mm}
\put(0,0){\circle{18}}
\put(-15,15){1}
\circulo{60}
\put(50,15){2}
\circulo{120}
\put(110,15){3}
\put(180,0){\circle{18}}
\put(170,15){4}
\linethickness{\thinn}
\qbezier(8,8)(60,90)(112,8)
\qbezier(68,-8)(120,-90)(172,-8)
\linethickness{\thickn}
\put(9,0){\liner}
\put(129,0){\liner}
\put(69,0){\liner}
\end{picture}}

\newcommand{\pthreezeroone}{\begin{picture}(\xmaxthree,\ymax)(-\xx,\yy)
\setlength{\unitlength}{.1mm}
\put(0,0){\circle{18}}
\put(-15,15){1}
\circulo{60}
\put(50,15){2}
\circulo{120}
\put(110,15){3}
\put(180,0){\circle{18}}
\put(170,15){4}
\circulo{240}
\put(230,15){5}
\linethickness{\thinn}
\qbezier(8,8)(60,90)(112,8)
\qbezier(128,8)(180,90)(232,8)
\qbezier(68,-8)(120,-90)(172,-8)
\linethickness{\thickn}
\put(9,0){\liner}
\put(129,0){\liner}
\put(69,0){\liner}
\put(189,0){\liner}
\end{picture}}

\newcommand{\pthreeoneone}{\begin{picture}(\xmaxthree,\ymax)(-\xx,\yy)
\setlength{\unitlength}{.1mm}
\circulo{0}
\put(-15,15){$N$}
\put(60,0){\circle{18}}
\put(50,15){1}
\circulo{120}
\put(110,15){2}
\circulo{180}
\put(170,15){3}
\put(240,0){\circle{18}}
\put(230,15){4}
\circulo{300}
\put(290,15){5}
\linethickness{\thinn}
\qbezier(8,-8)(60,-90)(112,-8)
\qbezier(128,-8)(180,-90)(232,-8)
\qbezier(68,8)(120,90)(172,8)
\qbezier(188,8)(240,90)(292,8)
\linethickness{\thickn}
\put(9,0){\liner}
\put(129,0){\liner}
\put(69,0){\liner}
\put(189,0){\liner}
\put(249,0){\liner}
\end{picture}}

\newcommand{\poneexact}{\begin{picture}(\xmaxthree,\ymax)(-\xxb,\yy)
\setlength{\unitlength}{.1mm}
\circulo{0}
\put(-55,15){$N-1$}
\circulo{60}
\put(45,15){$N$}
\put(120,0){\circle{18}}
\put(110,15){1}
\put(180,0){\circle{18}}
\put(170,15){2}
\circulo{240}
\put(230,15){3}
\circulo{300}
\put(290,15){4}
\linethickness{\thinn}
\qbezier(-112,8)(-60,90)(-8,8)
\qbezier(-52,-8)(0,-90)(52,-8)
\qbezier(8,8)(60,90)(112,8)
\qbezier(128,8)(180,90)(232,8)
\qbezier(248,8)(300,90)(352,8)
\qbezier(68,-8)(120,-90)(172,-8)
\qbezier(188,-8)(240,-90)(292,-8)
\qbezier(308,-8)(360,-90)(412,-8)
\multiput(-9, 0)(-16,0){7}{\line(-1,0){10}}
\multiput(308, 0)(16,0){7}{\linerr}
\linethickness{\thickn}
\put(9,0){\liner}
\put(69,0){\liner}
\multiput(129, 0)(16,0){3}{\linerr}
\put(189,0){\liner}
\put(249,0){\liner}
\end{picture}}

\newcommand{\ptwoexact}{\begin{picture}(\xmaxthree,\ymax)(-\xxb,\yy)
\setlength{\unitlength}{.1mm}
\circulo{0}
\put(-55,15){$N-1$}
\circulo{60}
\put(45,15){$N$}
\put(120,0){\circle{18}}
\put(110,15){1}
\circulo{180}
\put(170,15){2}
\put(240,0){\circle{18}}
\put(230,15){3}
\circulo{300}
\put(290,15){4}
\linethickness{\thinn}
\qbezier(-112,8)(-60,90)(-8,8)
\qbezier(-52,-8)(0,-90)(52,-8)
\qbezier(8,8)(60,90)(112,8)
\qbezier[20](128,8)(180,90)(232,8)
\qbezier(248,8)(300,90)(352,8)
\qbezier(68,-8)(120,-90)(172,-8)
\qbezier(188,-8)(240,-90)(292,-8)
\qbezier(308,-8)(360,-90)(412,-8)
\multiput(-9, 0)(-16,0){7}{\line(-1,0){10}}
\multiput(308, 0)(16,0){7}{\linerr}
\linethickness{\thickn}
\put(9,0){\liner}
\put(69,0){\liner}
\put(129, 0){\liner}
\put(189,0){\liner}
\put(249,0){\liner}
\end{picture}}

\begin{document}

\title{One-Dimensional Fluids with Second Nearest--Neighbor Interactions
}


\author{Riccardo Fantoni         \and
        Andr\'es Santos
}


\institute{R. Fantoni \at
              Universit\`a di Trieste, Dipartimento di Fisica, strada  Costiera 11, 34151 Grignano (Trieste), Italy \\
                           \email{rfantoni@ts.infn.it}           
           \and
           A. Santos \at
              Departamento de F\'{\i}sica and Instituto de Computaci\'on Cient\'{\i}fica  Avanzada (ICCAEx), Universidad de Extremadura,  06006 Badajoz, Spain\\
                           \email{andres@unex.es}
}


\maketitle

\begin{abstract}
As is well known, one-dimensional systems with interactions restricted to first nearest neighbors admit a full analytically exact statistical-mechanical solution. This is essentially due to the fact that the knowledge of the first nearest--neighbor probability distribution function,  $p_1(r)$, is enough to determine the structural and thermodynamic properties of the system.
On the other hand, if the interaction between second nearest--neighbor particles is turned on, the analytically exact solution is lost. Not only the knowledge of $p_1(r)$ is not sufficient anymore, but even its determination becomes a complex many-body problem.  In this work we systematically explore different approximate solutions for one-dimensional second nearest--neighbor fluid models. We apply those approximations to the square-well and the attractive two-step pair potentials and compare them with Monte Carlo simulations, finding an excellent agreement.
\end{abstract}

\section{Introduction}
\label{sec:introduction}

It is well known that equilibrium systems confined in one-dimensional  geometries with interactions restricted to first nearest
neighbors (1st nn) admit a full exact statistical-mechanical solution \cite{R91,K91,HGM34,T36,N40a,N40b,T42,SZK53,K55b,LPZ62,KT68,LZ71,P76,P82,BOP87,HC04,S07,BNS09}. Apart from its
undoubtful pedagogical and illustrative values \cite{BB83a,B83b,BB83d,B89,S16,HB08,LNP_book_note_13_08,LNP_book_note_15_05_1,LNP_book_note_15_06_1}, this exact solution can also be exploited as a benchmark for approximations \cite{BB83b,BB83c,B84,BS86,BE02,S07b,S07,AE13,ACE17} or simulation methods \cite{BB74}.

Exact solutions are also possible for a few one-dimensional systems
with interactions extending beyond 1st nn, as happens, for
example, in the one- and two-component plasma, the Kac--Backer model, or isolated self-gravitating system
\cite{R71b,F16,F17}. However,  most one-dimensional non-1st nn fluids do not admit an analytical exact solution but only
an approximate one, as happens, for example, to the penetrable-square-well model \cite{SFG08,FGMS10,F10b}.

While a fair amount of simplification of relevant questions occurs in a lattice gas or Ising model context \cite{BOP87}, here we will be concerned with spatially continuous fluid systems.
The two main ingredients allowing for an exact statistical-mechanical treatment of one-dimensional fluids (in the isothermal-isobaric ensemble) are \cite{S16}: (i) the pair interaction potential diverges as the two particles approach each other, so that the ordering of the particles cannot change, and (ii) each particle interacts only with its two 1st nn.  In that case, it is possible to obtain the exact 1st nn probability distribution function,  $p_1(r)$, whose knowledge is in turn enough to determine all the structural and thermodynamic properties of the system.

On the other hand, even if the ordering property (i) is maintained, as soon as the interaction extends to second nearest neighbors (2nd
nn) the exact solution is generally lost.
First, the determination of $p_1(r)$ becomes a complex many-body problem. Second, even if $p_1(r)$ were known, the convolution property relating $p_1(r)$ to the more general correlation functions is no longer valid and one is again faced with a many-body correlation coupling.
Nonetheless, the
problem is in general more tractable than the one of a generic non-1st nn fluid since the potential energy now contains only the
interactions between 1st and 2nd nn pairs. It is
therefore interesting to find reasonable approximate solutions in this particular
case. This is the objective of the present work.

After revising the exact expressions for the $\ell$th-order nn distribution functions in the isothermal-isobaric ensemble and their structure, we devise a sequence of approximations (by means of a diagrammatic description) at various increasing orders of accuracy.
Our sequence gives the exact solution
only at infinite order, but we will discover that already at
second order it does a very good job. As illustrations, we will  apply our approach to two particular cases: the square-well (SW) and the attractive two-step (TS) models.

As for the thermodynamic properties, the equation of state is determined from  three alternative roots: the virial and compressibility routes, and the consistency condition that the radial distribution function (RDF) must tend to $1$ at large distances. This gives us useful thermodynamic consistency tests on our approximations.
Of course, the van Hove theorem \cite{vH50,H63,R99} states that in our case there
cannot be a phase transition for the fluid and, in particular, the
isothermal susceptibility cannot diverge. We check this by computing
the isothermal susceptibility through two different thermodynamic
routes. Another relevant thermodynamic consistency test refers to the internal energy per particle.

We carry on a detailed analysis of the RDF and compare the behavior of our approximations with the results from canonical Monte Carlo (MC) simulations for both the SW and TS models. Also, within our approximate theory, we compute the Fisher--Widom (FW) line \cite{FW69} for the SW model at various ranges.

The work is organized as follows. In  Sect.\ \ref{sec:problem} the problem of the 2nd nn fluid is presented and the exact solution in the 1st nn case is recalled. In Sect.\ \ref{sec:approx} we introduce the sequence of approximations
used to solve the 2nd nn problem. This is followed by  Sects.\ \ref{sec:SW} and \ref{sec:TS}, where the approximations are particularized to the SW and TS fluids, and compared with our own MC simulations. In  Sect.\ \ref{sec:WFl}
we calculate the FW line for the SW model. Finally,
Sect.\ \ref{sec:conclusions} presents our concluding remarks.

\section{The roblem}
\label{sec:problem}

\begin{figure}
\begin{center}
\includegraphics[width=.9\columnwidth]{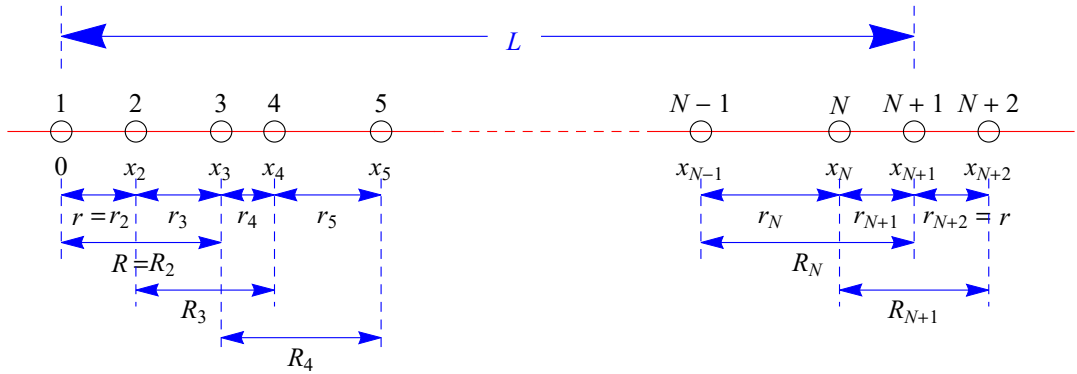}
\end{center}
\caption{Illustration of the one-dimensional fluid and of the
  distances used in the main text. For simplicity, particle $i=1$ defines the origin of coordinates (i.e., $x_1=0$).}
\label{fig:diag}
\end{figure}

Let us consider a one-dimensional system of $N$ particles in a box of
length $L$ (so that the number density is $n=N/L$) subject to a pair
interaction potential $\phi(r)$ such that:
\begin{itemize}
\item[i.] $\lim_{r\to 0}\phi(r)=\infty$. This implies that the   \emph{order} of the particles in the line does not change, i.e., the particles
  are assumed to be \emph{impenetrable}.
\item[ii.] $\phi(r)=0$ for $r>D$. Thus, the interaction has a \emph{finite  range} $D$.
\item[iii.] Each particle interacts \emph{only} with its 1st and 2nd
  nn, i.e., with the four particles closer to it.
\end{itemize}
The total potential energy is then
\bq \label{pp}
\Phi_N(\mathbf{r}^N)=\sum_{i=1}^N\left[\phi(x_{i+1}-x_i)+\gamma \phi(x_{i+2}-x_i)\right],
\eq
where $\mathbf{r}^N=\{x_1, x_2, \ldots ,x_N\}$ are the coordinates of the $N$ particles ordered in such way that $x_1<x_2< \cdots <x_N$, and periodic boundary conditions (pbc) are assumed, so that $x_{N+1}=x_1+L$ and $x_{N+2}=x_2+L$. A sketch of the system is shown in Fig.\ \ref{fig:diag}.
In Eq.\ \eqref{pp} we have introduced the bookkeeping factor $\gamma$ simply to keep track of the 2nd nn contribution to the total potential energy. At the end of the calculations $\gamma=1$ will be taken.

\subsection{General Relations}
\label{sec:nn}

\subsubsection{Nearest--Neighbor and Pair Correlation Functions}

Given a reference particle at a certain position, let $p_1(r)\rmd r$ be the \emph{conditional} probability of finding its 1st nn at a distance between $r$ and $r+\rmd r$ to its right.
More in general, we can define $p_\ell(r)\rmd r$ as the conditional probability of finding its
(right) $\ell$th neighbor ($1\leq\ell\leq N-1$) at a distance between
$r$ and $r+\rmd r$ \cite{S16}. Since the $\ell$th neighbor of the reference particle must be somewhere, the
normalization condition, in the thermodynamic limit ($N\to \infty$, $L\to \infty$, $n=\mathrm{const}$), is
\bq \label{norm}
\int_0^\infty \rmd r\,p_\ell(r)=1.
\eq

The physical
meaning of the RDF \cite{S16,H63,BH76,HM06} implies that
$ng(r)\rmd r$ is the \emph{total} number of particles at a distance between $r$
and $r+\rmd r$, regardless of whether those particles correspond to the
1st nn, the 2nd nn, $\ldots$ of the
reference particle. Thus, again in the thermodynamic limit,
\bq \label{RDF}
ng(r)=\sum_{\ell=1}^\infty p_\ell(r).
\eq

\subsubsection{Thermodynamic Quantities}
\label{subsubsec:thermo}

Apart from characterizing the equilibrium spatial correlations, the RDF allows one to obtain the thermodynamic quantities by means of well-known statistical-mechanical formulas \cite{S16}. For instance, the excess internal energy per particle is given by
\bq \label{er}
\beta u=\beta n\int_0^\infty \rmd r\,
\phi(r)g(r)=\beta \int_0^\infty \rmd r\,
\phi(r)\left[p_1(r)+p_2(r)\right],
\eq
where $\beta\equiv 1/k_BT$ is the (reduced) inverse temperature ($k_B$ and $T$ being the Boltzmann constant and the absolute temperature, respectively) and we have taken into account that $\phi(r)$ vanishes beyond 2nd nn.

Moreover, from the virial theorem  we find
\bq
\label{vr}
\frac{\beta p}{n}=1-\beta n \int_0^\infty \rmd r\,
r\phi'(r)g(r)
=1+n\int_0^\infty \rmd r\,rf'(r)y(r),
\eq
where $p$ is the pressure, $\cdots'\equiv \rmd\cdots/\rmd r$, $f(r)=\rme^{-\beta\phi(r)}-1$ is the
Mayer factor, and $y(r)=g(r)\rme^{\beta\phi(r)}$ is the cavity function.
A thermodynamic consistency test comes from the following
Maxwell relation
\bq \label{ct3}
\left(\frac{\partial\beta u}{\partial\beta p}\right)_\beta=
\beta\left(\frac{\partial n^{-1}}{\partial\beta}\right)_{\beta p}.
\eq
Of course,  for an exact solution this is an
identity.

The isothermal susceptibility is defined as
\bq \label{chia}
\chi_T=\left(\frac{\partial n}{\partial\beta p}\right)_\beta.
\eq
Alternatively, it can also be obtained via the compressibility route as \cite{S16}
\bq \label{chib}
\chi_T=1+2n\int_0^\infty \rmd r\, \left[g(r)-1\right]=1+2n\lim_{s\to 0}\left[\widehat{G}(s)-\frac{1}{s}\right],
\eq
where
\bq
\label{G(s)}
\widehat{G}(s)\equiv\int_0^\infty \rmd r\,\rme^{-rs}g(r)
\eq
is the Laplace transform of the RDF.
Again, the two routes (\ref{chia}) and
(\ref{chib})  give identical results if the exact RDF is used.
Note that the physical condition $\lim_{r\to\infty} g(r)=1$ and Eq.\ \eqref{chib}  imply the small-$s$ behavior
\bq
\label{small-s}
\widehat{G}(s)=\frac{1}{s}+\frac{\chi_T-1}{2n}+\Or(s).
\eq

\subsubsection{Isothermal-Isobaric Ensemble}

We will see that retaining in the potential energy of the fluid up to
the 2nd nn interactions involves an $N$-body coupling
in any of the $\ell$th nn distribution functions. This
renders the one-dimensional problem extremely more complicated than the
1st nn fluid, for which a general analytical solution
can be found \cite{S16} due to the decoupling of each pair of
nn (see  Sect.\ \ref{sec:1nn}).

In the isothermal-isobaric (or $NpT$) ensemble, the $N$-body configurational probability density function is \cite{S16}
\bq
\rho(\mathbf{r}^N)\propto \rme^{-\beta p L-\beta\Phi_N(\rr^N)}.
\eq
As a consequence,
we find for $p_1(r)$ (see Fig.\ \ref{fig:diag})
\bq
\label{p1(r)}
p_1(r)\propto\int_r^\infty \rmd L\,\rme^{-\beta pL}\int_{x_2}^L \rmd x_3
\int_{x_3}^L \rmd x_4\cdots\int_{x_{N-1}}^L \rmd x_N\,\rme^{-\beta\Phi_N(\rr^N)},
\eq
where we have taken particles $i=1$ (at $x_1=0$) and $i=2$ (at
$x_2=r$) as the representative 1st nn pair. The proportionality constant in Eq.\ \eqref{p1(r)} is obtained from the normalization condition  \eqref{norm}.

Using  Eq.\ (\ref{pp}) and the pbc we find (see Fig.\ \ref{fig:diag} for notation)
\beq
\label{PhiN}
\Phi_N(\rr^N)=\phi(r)+\gamma\phi(\rp)+\phi(r_3)+\gamma\phi(\rp_3)+\phi(r_4)+\gamma\phi(\rp_4)+
\cdots+\phi(r_{N+1})+\gamma\phi(\rp_{N+1}),
\eeq
with $\rp=x_3=r+r_3$ and $r_i=x_i-x_{i-1}$, $\rp_i=x_{i+1}-x_{i-1}=r_i+r_{i+1}$ for
$i=3,4,\ldots,N+1$.
It is proved in Appendix \ref{app:A} that, after an adequate change of variables, Eq.\ \eqref{p1(r)} becomes
\bq
\label{p1(r)2nn}
p_1(r)&\propto& \rme^{-\beta [pr+\phi(r)]}
\int_0^\infty \rmd r_3\,\rme^{-\beta[pr_3+\phi(r_3)]}\rme^{-\gamma\beta\phi(r+r_3)}
\int_0^{\infty}\rmd r_4\,\rme^{-\beta[pr_4+\phi(r_4)]}\rme^{-\gamma\beta\phi(r_3+r_4)}\cdots
 \nn
&&\times\int_0^{\infty}\rmd r_N\,\rme^{-\beta[pr_N+\phi(r_N)]}
\rme^{-\gamma\beta\phi(r_{N-1}+r_N)}\int_0^\infty \rmd r_{N+1}\rme^{-\beta[pr_{N+1}+\phi(r_{N+1})]}\rme^{-\gamma\beta\phi(r_N+r_{N+1})}\rme^{-\gamma\beta\phi(r_{N+1}+r)}.\nn
\eq
We clearly see  that all the $N-1$ spatial integrals are coupled, so that we cannot proceed any further without introducing approximations.
This many-body coupling can  be conveniently visualized by means of a diagrammatic representation, as shown in the first row of Table \ref{tab:diagrams}.

\begin{table}
\caption{Diagrammatic representation. We indicate with $(i,j)$ the distance
  $r_{ij}=|x_i-x_j|$. The open circles denote the root points (not
  integrated out) $i$ and $j$, while the filled circles (enclosed by open ones) denote field points (integrated
  out). The thick straight lines represent a link
  $\exp\{-\beta[pr+\phi(r)]\}$ between 1st nn, while the thin
  curved lines represent a link $\exp[-\gamma\beta \phi(r)]$ between
  2nd nn. In this representation, the dashed lines link two root points, while the solid lines link two field points or one root
  and one field point.}
\label{tab:diagrams}
\begin{center}
\begin{tabular}{lrll}
\hline\noalign{\smallskip}
Label&Function&&Diagram\\
\noalign{\smallskip}\hline\noalign{\smallskip}
exact&$p_1(1,2)$&$\propto$&$\poneexact$\\
exact&$p_2(1,3)$&$\propto$&$ \ptwoexact$\\
exact&$p_3(1,4)$&$\propto$&$ \pthreeexact$\\
\onezerozero&$p_1^{(00)}(1,2)$&$\propto$&$\ponezerozero$\\
\onezeroone&$p_1^{(01)}(1,2)$&$\propto$&$\ponezeroone=\ponezerozero\times\monezeroone$\\
\oneoneone&$p_1^{(11)}(1,2)$&$\propto$&$\poneoneone=\ponezerozero\times\Bigg(\monezeroone\Bigg)^2$\\
\twozerozero&$p_2^{(00)}(1,3)$&$\propto$&$\ptwozerozero=\eonethree\times\Bigg(\ponezerozero * \ponezerozerobis\Bigg)$\\
\twozeroone&$p_2^{(01)}(1,3)$&$\propto$&$ \ptwozeroone=\eonethree\times \Bigg(\ponezerozero * \ponezeroonebis\Bigg)$\\
\twooneone&$p_2^{(11)}(1,3)$&$\propto$&$ \ptwooneone=\eonethree\times \Bigg(\ponezeroonebisbis * \ponezeroonebis\Bigg)$\\
\threezerozero&$p_3^{(00)}(1,4)$&$\propto$&$ \pthreezerozero$ \\
\threezeroone&$p_3^{(01)}(1,4)$&$\propto$&$ \pthreezeroone$\\
\threeoneone&$p_3^{(11)}(1,4)$&$\propto$&$\pthreeoneone$\\
&&&\\
\noalign{\smallskip}\hline
\end{tabular}
\end{center}
\end{table}

For the 2nd nn distribution we have
\bq
\label{p2(r)}
p_2(\rp)\propto\int_{\rp}^\infty \rmd L\,\rme^{-\beta p L}
\int_0^{\rp} \rmd x_2\int_{\rp}^L \rmd x_4\int_{x_4}^L \rmd x_5\cdots
\int_{x_{N-1}}^L \rmd x_N\,\rme^{-\beta\Phi_N(\rr^N)}.
\eq
As proved in  Appendix \ref{app:A}, this becomes
\bq
\label{p2exact}
p_2(\rp)&\propto&\rme^{-\beta [p \rp+\gamma\phi(\rp)]}\int_0^{\rp}
\rmd r_2\,\rme^{-\beta\phi(r_2)}\rme^{-\beta\phi(\rp-r_2)}\int_0^\infty \rmd r_4\,\rme^{-\beta[pr_4+\phi(r_4)]}
\rme^{-\gamma\beta\phi(\rp-r_2+r_4)}\nn
&&\times
\int_0^\infty
\rmd r_5\,\rme^{-\beta[pr_5+\phi(r_5)]}\rme^{-\gamma\beta\phi(r_4+r_5)}\cdots
\int_0^\infty
\rmd r_N\,\rme^{-\beta[pr_N+\phi(r_N)]}\rme^{-\beta\phi(r_{N-1}+r_N)}\nn
&&\times \int_0^\infty \rmd r_{N+1}\,\rme^{-\beta[pr_{N+1}+\phi(r_{N+1})]}
\rme^{-\gamma\beta\phi(r_N+r_{N+1})}\rme^{-\gamma\beta\phi(r_{N+1}+r_2)}.
\eq
Once again, the expression above depends on all the $N$-body
terms. It is represented by the second row in   Table \ref{tab:diagrams}.

In the case of the 3rd nn distribution, its formal expression is
\bq
\label{p3(r)}
p_3(\rpp)\propto\int_{\rpp}^\infty \rmd L\,\rme^{-\beta pL}
\int_0^{\rpp} \rmd x_2\int_{x_2}^{\rpp} \rmd x_3\int_{\rpp}^L \rmd x_5\cdots
\int_{x_{N-1}}^L \rmd x_N\,\rme^{-\beta\Phi_N(\rr^N)},
\eq
where we have denoted by $\rpp$ the distance between the reference particles $1$ and $4$.
Equation \eqref{p3(r)} is equivalent to (see Appendix \ref{app:A})
\bq
\label{p3exact}
p_3(\rpp)&\propto&\rme^{-\beta p \rpp}\int_0^{\rpp}\rmd r_2\,\rme^{-\beta\phi(r_2)}\rme^{-\gamma\beta\phi(\rpp-r_2)}
\int_0^{\rpp-r_2}\rmd r_3\,\rme^{-\beta\phi(r_3)}\rme^{-\gamma\beta\phi(r_2+r_3)}\rme^{-\beta\phi(\rpp-r_2-r_3)}\nn
&&\times\int_0^\infty \rmd r_5\,\rme^{-\beta[pr_5+\phi(r_5)]}\rme^{-\gamma\beta\phi(\rpp-r_2-r_3+r_5)}
\int_0^\infty
\rmd r_6\,\rme^{-\beta[pr_6+\phi(r_6)]}\rme^{-\gamma\beta\phi(r_5+r_6)}\cdots \nn
&&\times\int_0^\infty
\rmd r_N\,\rme^{-\beta[pr_N+\phi(r_N)]}\rme^{-\gamma\beta\phi(r_{N-1}+r_N)}\int_0^\infty \rmd r_{N+1}\,\rme^{-\beta[pr_{N+1}+\phi(r_{N+1})]}
\rme^{-\gamma\beta\phi(r_N+r_{N+1})}\rme^{-\gamma\beta\phi(r_{N+1}+r_2)}.\nn
\eq
The diagram representing Eq.\ \eqref{p3exact} is displayed as the third row of Table \ref{tab:diagrams}.

The process can be continued in a similar way to get $p_{\ell}(\rpp)$ with $\ell=4,\ldots,N-1$ and $\rpp=x_{\ell+1}-x_1$. In particular,
\bq
\label{pN-1exact}
p_{N-1}(\rpp)&\propto&\rme^{-\beta p \rpp}\int_0^{\rpp} \rmd r_2\,\rme^{-\beta\phi(r_2)}\int_0^{\rpp-r_2} \rmd r_3\,\rme^{-\beta\phi(r_3)}\rme^{-\gamma\beta\phi(r_2+r_3)}\int_0^{\rpp-r_2-r_3}\rmd r_4\,\rme^{-\beta\phi(r_4)}\rme^{-\gamma\beta\phi(r_3+r_4)}
\cdots
\nn
&&\times
\int_0^{\rpp-r_2-r_3-\cdots-r_{N-2}}\rmd r_{N-1}
\rme^{-\beta\phi(r_{N-1})}\rme^{-\gamma\beta\phi(r_{N-2}+r_{N-1})}
\rme^{-\beta\phi(\rpp-r_2-r_3-\cdots-r_{N-1})}
\rme^{-\gamma\beta\phi(\rpp-r_2-r_3-\cdots-r_{N-2})}\nn
&&\times\int_0^\infty \rmd r_{N+1}\,
\rme^{-\beta[p r_{N+1}+\phi(r_{N+1})]}
\rme^{-\gamma\beta\phi(\rpp-r_2-r_3-\cdots-r_{N-1}+r_{N+1})}
\rme^{-\gamma\beta\phi(r_{N+1}+r_2)}.
\eq

\subsection{First Nearest--Neighbor Fluids: Exact Solution}
\label{sec:1nn}
Let us suppose now that the 2nd nn interactions are switched off. This is equivalent to setting $\gamma=0$ in Eqs.\ \eqref{pp} and \eqref{PhiN}. In that case, the curved lines in the three first rows of Table \ref{tab:diagrams} disappear and most of the integrals in Eqs.\ \eqref{p1(r)2nn}, \eqref{p2exact}, and \eqref{p3exact} can be absorbed into the proportionality constants:
\begin{subequations}
\label{p123_1nn}
\bq
\label{p1_1nn}
p_1(r)=K_1 \rme^{-\beta[pr+\phi(r)]},
\eq
\bq
\label{p2_1nn}
p_2(\rp)\propto\rme^{-\beta p \rp}\int_0^{\rp}
\rmd r_2\,\rme^{-\beta\phi(r_2)}\rme^{-\beta\phi(\rp-r_2)},
\eq
\bq
\label{p3_1nn}
p_3(\rpp)\propto\rme^{-\beta p \rpp}\int_0^{\rpp}\rmd r_2\,\rme^{-\beta\phi(r_2)}
\int_0^{\rpp-r_2}\rmd r_3\,\rme^{-\beta\phi(r_3)}\rme^{-\beta\phi(\rpp-r_2-r_3)},
\eq
\end{subequations}
where in Eq.\ \eqref{p1_1nn} $K_1$ is the normalization constant.
In the case of Eq.\ \eqref{pN-1exact}, even though only the integral over $r_{N+1}$ can be absorbed into the proportionality constant so that $N-2$ integrals still remain, they acquire a simple convolution structure:
\bq
\label{pN-1_1nn}
p_{N-1}(\rpp)&\propto&\rme^{-\beta p \rpp}\int_0^{\rpp} \rmd r_2\,\rme^{-\beta\phi(r_2)}\int_0^{\rpp-r_2} \rmd r_3\,\rme^{-\beta\phi(r_3)}
 \int_0^{\rpp-r_2-r_3}\rmd r_4\,\rme^{-\beta\phi(r_4)}\cdots\nn
&&\times
\int_0^{\rpp-r_2-r_3-\cdots-r_{N-2}}\rmd r_{N-1}
\rme^{-\beta\phi(r_{N-1})}
\rme^{-\beta\phi(\rpp-r_2-r_3-\cdots-r_{N-1})}.
\eq

Thus, in the case of a  pure 1st nn fluid,  the following recurrence
relation holds
\bq \label{conv}
p_\ell(r)=\int_0^r \rmd r'\,p_1(r')p_{\ell-1}(r-r')\equiv (p_1*p_{\ell-1})(r).
\eq
It is straightforward to check that Eqs.\ \eqref{p123_1nn} and \eqref{pN-1_1nn} are consistent with Eq.\ \eqref{conv}.
The convolution structure of the integral in Eq.\ \eqref{conv} suggests the introduction of the
Laplace transform
\bq
\widehat{p}_\ell(s)\equiv\int_0^\infty \rmd r\,\rme^{-rs}p_\ell(r),
\eq
so that  Eq.\ (\ref{conv}) becomes
\bq \label{convs}
\widehat{p}_\ell(s)=\widehat{p}_1(s)\widehat{p}_{\ell-1}(s)=\left[\widehat{p}_1(s)\right]^\ell.
\eq
The normalization condition \eqref{norm} is equivalent to
\bq
\widehat{p}_\ell(0)=1.
\eq
Note that this condition is automatically satisfied by Eq.\ \eqref{convs} provided that $\widehat{p}_1(0)=1$. In fact, the Laplace transform of Eq.\ \eqref{p1_1nn} is
\bq \label{1dnn}
\widehat{p}_1(s)=K_1\widehat{\Omega}(s+\beta p),\quad K_1=\frac{1}{\widehat{\Omega}(\beta p)},
\eq
where
\bq
\widehat{\Omega}(s)\equiv\int_0^\infty \rmd r\,\rme^{-sr}\rme^{-\beta\phi(r)}
\eq
is the Laplace transform of the pair Boltzmann factor $\rme^{-\beta\phi(r)}$.

In this case of a 1st nn fluid, the RDF in Laplace space is exactly given by [see Eqs.\ \eqref{RDF} and \eqref{G(s)}]
\bq \label{Gs1}
\widehat{G}(s)=\frac{1}{n}\sum_{\ell=1}^\infty
\left[\widehat{p}_1(s)\right]^\ell=\frac{1}{n}
\frac{\widehat{p}_1(s)}{1-\widehat{p}_1(s)}.
\eq
Finally, the number density $n$ is obtained as a function of pressure and temperature by enforcing the condition $\lim_{s\to 0}sG(s)=1$ [see Eq.\ \eqref{small-s}]. The result is \cite{S16}
 \bq
    n=-\frac{\widehat{\Omega}(\beta p)}{\widehat{\Omega}'(\beta p)},
    \label{6.15}
    \eq
where
\bq
\widehat{\Omega}'(s)\equiv \frac{\partial \widehat{\Omega}(s)}{\partial s}=-\int_0^\infty \rmd r\, \rme^{-rs}r\rme^{-\beta\phi(r)}.
\eq

Obviously, Eqs.\ \eqref{p123_1nn}--\eqref{conv}, \eqref{convs}, \eqref{1dnn}, \eqref{Gs1}, and \eqref{6.15} cease  to be exactly valid as soon as the interactions extend to 2nd nn (i.e., $\gamma=1$).

\section{Our Approximations}
\label{sec:approx}

\subsection{First Nearest--Neighbor Distribution}
\label{sec3.1}

As discussed above,  the exact expression (\ref{p1(r)2nn}) for $p_1$ is not amenable for an analytical treatment of the problem. We will then
introduce a hierarchy of successive approximations.

In Eq.\ \eqref{p1(r)2nn} we observe that, apart from the prefactor $\exp\{-\beta[pr+\phi(r)]\}$, the distance $r$ appears explicitly  in the first integral (over $r_3$) and, because of the pbc, in the last integral (over $r_{N+1}$). The dependence on $r$  propagates as well  to the remaining integrals (over  $r_4$, \ldots, $r_N$) due to the nested structure of the integrals induced by the 2nd nn  terms of the form $\exp[-\gamma\beta\phi(r_i+r_{i+1})]$, as diagrammatically illustrated in the first row of Table \ref{tab:diagrams}.
On the other hand, the $r$-dependence becomes more and more indirect and attenuated as the integrals involve particles farther and farther from the pair $(1,2)$, either to its right or (because of the pbc) to its left. Thus, by truncating the integrals at a certain order and incorporating their values into the normalization constant, one can construct a hierarchy of approximations to $p_1(r)$ involving only a finite number of particles in the environment of the pair $(1,2)$.

The crudest approximation would consist in just neglecting the $r$-dependence in all the integrals of Eq.\ \eqref{p1(r)2nn}, i.e.,
\bq
\label{(00)_1}
p_1^{(00)}(r)=K_1^{(00)}\rme^{-\beta[pr+\phi(r)]},
\eq
where $K_1^{(00)}$ is the normalization constant.  Henceforth, a factor of the form $K_{\ell}^{(k_1k_2)}$ will denote a normalization constant.
In the zeroth-order approximation \eqref{(00)_1}, represented by the diagram with the label {\onezerozero}
in Table \ref{tab:diagrams}, $p_1(r)$ is assumed to be given by  the exact solution \eqref{p1_1nn} for the 1st nn fluid.  It can reasonably be expected that this is a very poor approximation for the 2nd nn fluid.

A less trivial approximation is obtained by including the integral over $r_3$ but not the other ones, i.e.,
\bq
p_1^{(01)}(r)=K_1^{(01)}\rme^{-\beta[pr+\phi(r)]}
\int_0^\infty \rmd r_3\,\rme^{-\beta[p r_3+\phi(r_3)]}\rme^{-\beta\phi(r+r_3)},
\eq
where henceforth  $\gamma=1$ is already set. This first-order approximation to the exact $p_1(r)$ is represented by the diagram with the label {\onezeroone} in Table \ref{tab:diagrams}. If, instead of including the integral over $r_3$ (i.e., the distance between the root particle $2$ and the particle to its right) we include the integral over $r_{N+1}$ (i.e., the distance between the root particle $1$ and the particle to its left, according to the pbc) we have
\bq
p_1^{(10)}(r)=K_1^{(10)}\rme^{-\beta[pr+\phi(r)]}
\int_0^\infty \rmd r_{N+1}\,\rme^{-\beta[p r_{N+1}+\phi(r_{N+1}]}\rme^{-\beta\phi(r_{N+1}+r)}.
\eq
Since $r_3$ and $r_{N+1}$ are dummy integration variables, it is obvious that $p_1^{(10)}(r)=p_1^{(01)}(r)$, as expected by symmetry arguments.

The first-order approximation $p_1^{(01)}(r)$, while more reliable than $p_1^{(00)}(r)$, is asymmetric as it treats one side of the pair $(1,2)$ differently from the other side. This is remedied by the second-order approximation
\bq \label{(11)_1}
p_1^{(11)}(r)=K_1^{(11)}\rme^{-\beta[pr+\phi(r)]}
\left[\int_0^\infty \rmd r_3\,\rme^{-\beta[p r_3+\phi(r_3)]}\rme^{-\beta\phi(r+r_3)}\right]^2,
\eq
where we have exploited the fact that the integrals over $r_3$ and over $r_{N+1}$ are identical.
A diagram for this approximation is shown with the label {\oneoneone} in
 Table \ref{tab:diagrams}.

Obviously, the same scheme could be followed by introducing the approximations $p_1^{(12)}$, $p_1^{(22)}$, $p_1^{(33)}$, and so on. They become increasingly more accurate at the expense of becoming increasingly more involved. In fact, the exact 1st nn distribution is recovered, in the thermodynamic limit, as $p_1=\lim_{k\to\infty}p_1^{(kk)}$. As a compromise between accuracy and simplicity we stop at the second-order approximation $p_1^{(11)}$.

\subsection{Second Nearest--Neighbor Distribution}
\label{sec3.2}
A similar process can be followed for the 2nd nn distribution $p_2(\rp)$. Here, particles $1$ and $3$ are fixed and one needs to integrate over all the positions of the intermediate particle $2$. If one ignores in Eq.\ \eqref{p2exact} the $R$-dependence of the integrals over those field particles to the right of $3$ or to the left of $1$, one finds
\bq
\label{(00)_2}
p_2^{(00)}(\rp)&=&K_2^{(00)}\rme^{-\beta[p \rp+\phi(\rp)]}\int_0^{\rp}
\rmd r_2\,\rme^{-\beta\phi(r_2)}\rme^{-\beta\phi(\rp-r_2)}\nn
&=&\frac{K_2^{(00)}}{\left[K_1^{(00)}\right]^2}\rme^{-\beta\phi(\rp)}\int_0^{\rp} \rmd r_2\,p_1^{(00)}(r_2)p_1^{(00)}(\rp-r_2).
\eq
A diagram for this zeroth-order approximation is shown with the label {\twozerozero} in
 Table \ref{tab:diagrams}, where we have used the fact that the point $2$
is an articulation point to simplify the diagram as the
convolution of two sub-diagrams.

The asymmetric first-order  approximation for $p_2$ is
\bq
\label{(01)_2}
p_2^{(01)}(\rp)&=&K_2^{(01)}\rme^{-\beta[p \rp+\phi(\rp)]}\int_0^{\rp}
\rmd r_2\,\rme^{-\beta\phi(r_2)}\rme^{-\beta\phi(\rp-r_2)}\int_0^\infty \rmd r_4 \,\rme^{-\beta [p r_4+\phi(r_4)]}\rme^{-\beta\phi(R-r_2+r_4)}\nn
&=&\frac{K_2^{(01)}}{K_1^{(00)}K_1^{(01)}}\rme^{-\beta\phi(\rp)}\int_0^{\rp} \rmd r_2\,p_1^{(00)}(r_2)p_1^{(01)}(\rp-r_2).
\eq
This approximation is
described by the
diagram with the label {\twozeroone} in Table \ref{tab:diagrams},  where again the convolution property is used.

The symmetrization of $p_2^{(01)}$ gives rise to the second-order approximation
\bq
\label{(11)_2}
p_2^{(11)}(\rp)&=&K_2^{(11)}\rme^{-\beta[p \rp+\phi(\rp)]}\int_0^{\rp}
\rmd r_2\,\rme^{-\beta\phi(r_2)}\rme^{-\beta\phi(\rp-r_2)}\int_0^\infty \rmd r_4 \,\rme^{-\beta [p r_4+\phi(r_4)]}\rme^{-\beta\phi(R-r_2+r_4)}\nn
&&\times
\int_0^\infty \rmd r_{N+1}\,\rme^{-\beta[pr_{N+1}+\phi(r_{N+1})]}
\rme^{-\beta\phi(r_{N+1}+r_2)}.
\nn
&=&\frac{K_2^{(11)}}{\left[K_1^{(01)}\right]^2}\rme^{-\beta\phi(\rp)}\int_0^{\rp} \rmd r_2\,p_1^{(01)}(r_2)p_1^{(01)}(\rp-r_2).
\eq
A diagram for this approximation is shown with label {\twooneone} in Table \ref{tab:diagrams}. Again,  the point $2$
is an articulation point so that the diagram simplifies as the
convolution of two sub-diagrams.

As in the case of $p_1$, one could define $p_2^{(22)}$, $p_2^{(33)}$, \ldots, but for simplicity we stop at the level of the {\twooneone} approximation \eqref{(11)_2}.

\subsection{Third Nearest--Neighbor Distribution}
\label{sec3.3}

Regarding the 3rd nn probability distribution, we can proceed by starting from Eq.\ \eqref{p3exact} and introducing the zeroth-, first-, and second-order approximations. They are given by
\beq
\label{000}
p_3^{(00)}(\rpp)=K_3^{(00)}\rme^{-\beta p \rpp}\int_0^{\rpp}\rmd r_2\,\rme^{-\beta\phi(r_2)}\rme^{-\beta\phi(\rpp-r_2)}
\int_0^{\rpp-r_2}\rmd r_3\,\rme^{-\beta\phi(r_3)}\rme^{-\beta\phi(r_2+r_3)}
\rme^{-\beta\phi(\rpp-r_2-r_3)},
\eeq
\bq
\label{110}
p_3^{(01)}(\rpp)&=&K_3^{(01)}\rme^{-\beta p \rpp}\int_0^{\rpp}\rmd r_2\,\rme^{-\beta\phi(r_2)}\rme^{-\beta\phi(\rpp-r_2)}
\int_0^{\rpp-r_2}\rmd r_3\,\rme^{-\beta\phi(r_3)}\rme^{-\beta\phi(r_2+r_3)}\rme^{-\beta\phi(\rpp-r_2-r_3)}\nn
&&\times\int_0^\infty \rmd r_5\,\rme^{-\beta[pr_5+\phi(r_5)]}\rme^{-\beta\phi(\rpp-r_2-r_3+r_5)},
\eq
\bq
\label{111}
p_3^{(11)}(\rpp)&=&K_3^{(11)}\rme^{-\beta p \rpp}\int_0^{\rpp}\rmd r_2\,\rme^{-\beta\phi(r_2)}\rme^{-\beta\phi(\rpp-r_2)}
\int_0^{\rpp-r_2}\rmd r_3\,\rme^{-\beta\phi(r_3)}\rme^{-\beta\phi(r_2+r_3)}\rme^{-\beta\phi(\rpp-r_2-r_3)}\nn
&&\times\int_0^\infty \rmd r_5\,\rme^{-\beta[pr_5+\phi(r_5)]}\rme^{-\beta\phi(\rpp-r_2-r_3+r_5)}
 \int_0^\infty \rmd r_{N+1}\,\rme^{-\beta[pr_{N+1}+\phi(r_{N+1})]}\rme^{-\beta\phi(r_{N+1}+r_2)}.
\eq
These approximations are represented by the diagrams labeled {\threezerozero}, {\threezeroone}, and {\threeoneone}, respectively, in Table \ref{tab:diagrams}. Since there are no articulation points, the diagrams cannot be simplified any further.

By following the same process one could construct similar approximations for $p_4$, $p_5$, \ldots, but they become increasingly more intricate as they would involve at least three, four, \ldots \, nested integrals.

\subsection{Radial Distribution Function}
\label{sec3.4}

As clearly seen from Eq.\ \eqref{RDF}, the knowledge (even if it were exact) of $p_1(r)$, $p_2(r)$, and $p_3(r)$ is not enough to get the RDF $g(r)$, as we need $p_\ell(r)$ for $\ell\geq 4$ as well. Thus, additional approximations are required.

Assume first that we want to construct an approximate function $\widehat{G}(s)$ based on $\widehat{p}_1(s)$ and  $\widehat{p}_2(s)$ only (since it is essential to keep at least those two  quantities in a 2nd nn fluid). How can we estimate $\widehat{p}_\ell (s)$ with $\ell\geq 3$ from $\widehat{p}_1(s)$ and  $\widehat{p}_2(s)$? A simple possibility consists in extending the exact convolution property \eqref{convs} of 1st nn fluids as an approximation to 2nd nn fluids. Two main possibilities arise:
\begin{subequations}
\bq
\label{p1p2a}
\widehat{p}_{2\ell+1}(s)=\widehat{p}_1(s)\left[\widehat{p}_2(s)\right]^\ell,\quad \widehat{p}_{2\ell+2}(s)=\left[\widehat{p}_2(s)\right]^{\ell+1}, \quad \ell\geq 1,
\eq
\bq
\label{plp2b}
\widehat{p}_{\ell}(s)=\left[\widehat{p}_1(s)\right]^\ell,\quad  \quad \ell\geq 3.
\eq
\end{subequations}
Then, application of Eq.\ \eqref{RDF} yields, respectively,
\begin{subequations}
\label{RDF2ab}
\bq \label{RDF2}
\widehat{G}(s)=\frac{1}{n}\frac{\widehat{p}_1(s)+\widehat{p}_2(s)}
{1-\widehat{p}_2(s)},\quad {n}=\frac{2}{\widetilde{p}_2},
\eq
\bq \label{RDF2b}
\widehat{G}(s)=\frac{1}{n}\left\{
\frac{\widehat{p}_1(s)}
{1-\widehat{p}_1(s)}+\widehat{p}_2(s)-\left[\widehat{p}_1(s)\right]^2\right\}
,\quad {n}=\frac{1}{\widetilde{p}_1}.
\eq
\end{subequations}
Here, we have used the condition $\lim_{s\to 0} s\widehat{G}(s)=1$ [see Eq.\ \eqref{small-s}] to determine the number density $n$ in terms of
\bq
\widetilde{p}_\ell\equiv -\left.\frac{\partial \widehat{p}_\ell(s)}{\partial s}\right|_{s=0}=\int_0^\infty \rmd r\, rp_\ell (r).
\eq
Regardless of whether Eqs.\ \eqref{RDF2} or  \eqref{RDF2b} is used, a different approximation for $\widehat{G}(s)$ is made depending on which approximation is chosen for $p_1$ (see  Sect.\ \ref{sec3.1}) and $p_2$ (see Sect.\ \ref{sec3.2}). We introduce the notation $[1^{(\alpha_1)}2^{(\alpha_2)}]_a$ and $[1^{(\alpha_1)}2^{(\alpha_2)}]_b$ to refer to Eqs.\ \eqref{RDF2} and \eqref{RDF2b}, respectively,  complemented with the approximations $1^{(\alpha_1)}$  for $p_1$ and $2^{(\alpha_2)}$ for $p_2$,  where $(\alpha_1),(\alpha_2)=(00), (01)$, or $(11)$.

In Eqs.\ \eqref{RDF2ab}  $\widehat{p_3}(s)$ is expressed in terms of $\widehat{p}_1(s)$ and $\widehat{p}_2(s)$. On the other hand, if the 3rd nn probability distribution is described, with independence of $\widehat{p}_1(s)$ and $\widehat{p}_2(s)$, by any of the approximation of  Sect.\ \ref{sec3.3}  we can construct $\widehat{p}_\ell(s)$ with $\ell\geq 4$ as any of the following three possibilities:
\begin{subequations}
\beq
\label{p1p2p3a}
\widehat{p}_{3\ell+1}(s)=\widehat{p}_1(s)\left[\widehat{p}_3(s)\right]^\ell,\quad \widehat{p}_{3\ell+2}(s)=\widehat{p}_2(s)\left[\widehat{p}_3(s)\right]^\ell,\quad
\widehat{p}_{3\ell+3}(s)=\left[\widehat{p}_3(s)\right]^{\ell+1}, \quad \ell\geq 1,
\eeq
\bq
\label{p1p2p3b}
\widehat{p}_{2\ell}(s)=\left[\widehat{p}_2(s)\right]^{\ell}, \quad \widehat{p}_{2\ell+1}(s)=\widehat{p}_1(s)\left[\widehat{p}_2(s)\right]^{\ell}, \quad \ell\geq 2,
\eq
\bq
\label{plp2p3c}
\widehat{p}_{\ell}(s)=\left[\widehat{p}_1(s)\right]^\ell,\quad  \quad \ell\geq 4.
\eq
\end{subequations}
This gives rise, respectively, to
\begin{subequations}
\label{RDF3abc}
\bq \label{RDF3a}
\widehat{G}(s)=\frac{1}{n}\frac{\widehat{p}_1(s)+\widehat{p}_2(s)+\widehat{p}_3(s)}
{1-\widehat{p}_3(s)},\quad {n}=\frac{3}{\widetilde{p}_3},
\eq
\bq \label{RDF3b}
\widehat{G}(s)=\frac{1}{n}\left\{
\frac{\widehat{p}_1(s)+\widehat{p}_2(s)}
{1-\widehat{p}_2(s)}+\widehat{p}_3(s)-
\widehat{p}_1(s)\widehat{p}_2(s)\right\},\quad {n}=\frac{2}{\widetilde{p}_2},
\eq
\bq
\label{RDF3c}
\widehat{G}(s)=\frac{1}{n}\left\{
\frac{\widehat{p}_1(s)}
{1-\widehat{p}_1(s)}+\widehat{p}_2(s)+\widehat{p}_3(s)-
\left[\widehat{p}_1(s)\right]^2-
\left[\widehat{p}_1(s)\right]^3\right\},\quad {n}=\frac{1}{\widetilde{p}_1}.
\eq
\end{subequations}
As before, we will denote as $[1^{(\alpha_1)}2^{(\alpha_2)}3^{(\alpha_3)}]_a$, $[1^{(\alpha_1)}2^{(\alpha_2)}3^{(\alpha_3)}]_b$, and $[1^{(\alpha_1)}2^{(\alpha_2)}3^{(\alpha_3)}]_c$ the approximations \eqref{RDF3a}, \eqref{RDF3b}, and \eqref{RDF3c}, respectively, complemented with the approximations $1^{(\alpha_1)}$ for $p_1$, $2^{(\alpha_2)}$  for $p_2$, and $3^{(\alpha_3)}$  for $p_3$.

Note that Eqs.\ \eqref{RDF2ab} and \eqref{RDF3abc} are fully equivalent in the case of a 1st nn fluid, as a consequence of Eq.\ \eqref{convs}. This is not so, however, for 2nd nn fluids. On physical grounds, the approximations of the form $[1^{(\alpha_1)}2^{(\alpha_2)}]_a$, where $\widehat{p}_3(s)\approx \widehat{p}_1(s)\widehat{p}_2(s)$, are expected to be more accurate than those of the form $[1^{(\alpha_1)}2^{(\alpha_2)}]_b$, where $\widehat{p}_3(s)\approx \left[\widehat{p}_1(s)\right]^3$. Likewise, the approximations of the form $[1^{(\alpha_1)}2^{(\alpha_2)}3^{(\alpha_3)}]_a$ are expected to be better than those of the form $[1^{(\alpha_1)}2^{(\alpha_2)}3^{(\alpha_3)}]_b$ or, even more, of the form $[1^{(\alpha_1)}2^{(\alpha_2)}3^{(\alpha_3)}]_c$.

We will now apply our approximations to two specific 2nd nn fluid models and assess the results by MC simulations.

\section{The Square-Well Potential}
\label{sec:SW}
As a simple prototype potential to test our approach, let us consider the SW potential,
\bq \label{SWpotential}
\phi(r)=\left\{\begin{array}{ll}
\infty, & r<\sigma,\\
-\epsilon, & \sigma\leq r<\lambda\sigma,\\
0, & \lambda\sigma\leq r.
\end{array}\right.
\eq
The physical properties of the fluid will
depend on the dimensionless range $\lambda$, the
reduced temperature $T^*\equiv k_BT/\epsilon$, and either the reduced density
$n^*\equiv n\sigma$ or the reduced pressure  $p^*\equiv p\sigma/\epsilon$.
Of course, for $\lambda\leq 2$ the fluid is a 1st nn
one so it admits an exact solution \cite{KT68,S16,LNP_book_note_15_05_1}. Our results for the 2nd
nn fluid will allow us to extend such an exact solution, in
an approximate way, to the range $2<\lambda\leq 3$.

\subsection{Structural Properties}

Clearly, due to the hard core at $r=\sigma$, one has $p_\ell(r)=0$ for $r<\ell\sigma$. Therefore,
\bq
\label{4.2}
ng(r)=\left\{
\begin{array}{ll}
 p_1(r), &\sigma< r<2\sigma,\\
  p_1(r)+p_2(r),&2\sigma<r<3\sigma,\\
  p_1(r)+p_2(r)+p_3(r),&3\sigma<r<4\sigma,\\
  \vdots&\vdots
\end{array}
\right.
\eq

All the approximations for the 1st nn, 2nd nn, and 3rd nn distributions described in Sects.\ \ref{sec3.1}--\ref{sec3.3} have fully analytical (albeit too long to be displayed here) expressions in the case of the SW potential, both in real space and in Laplace space. This allows one to obtain analytical expressions for the Laplace transform $\widehat{G}(s)$ in any of the approximations described in Sect.\ \ref{sec3.4}. The RDF in real space, $g(r)$, can then be found up to $r=4\sigma$ by application of Eq.\ \eqref{4.2} and, for longer distances, by a numerical inverse Laplace transform using the algorithm described in  Ref.\ \cite{AW92}.

Henceforth, unless stated otherwise, we take $\sigma=1$ as the length unit and particularize to $\lambda=3$, which is the largest range consistent with 2nd nn interactions.

\begin{figure}
\begin{center}
\includegraphics[width=0.45\columnwidth]{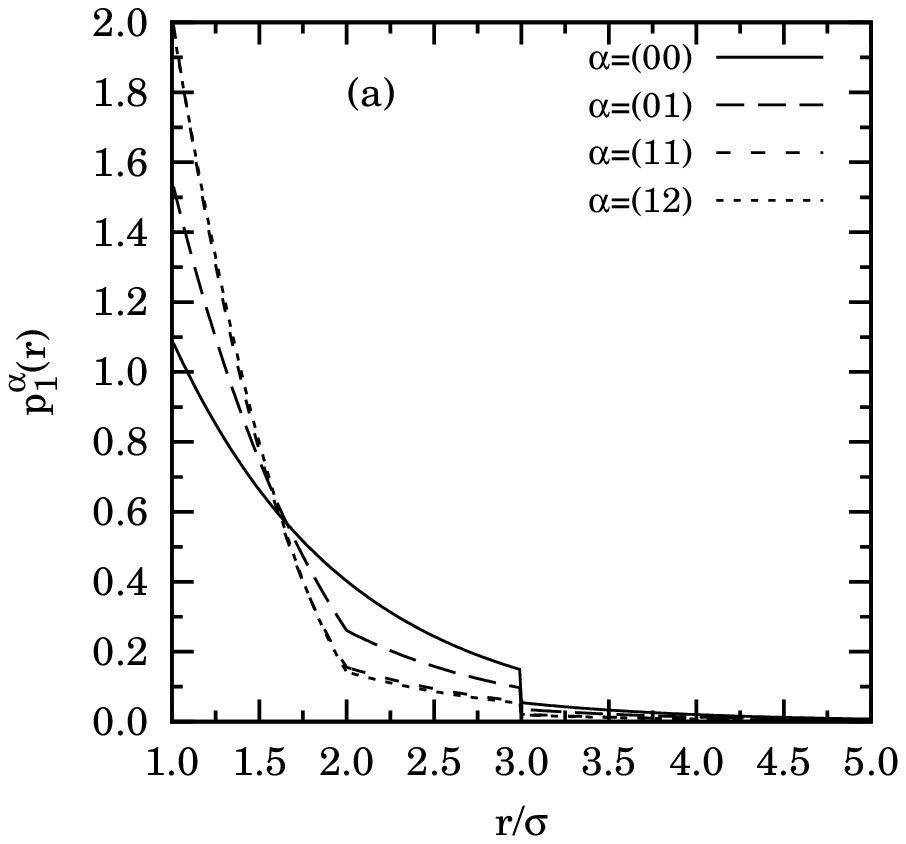}
\includegraphics[width=0.45\columnwidth]{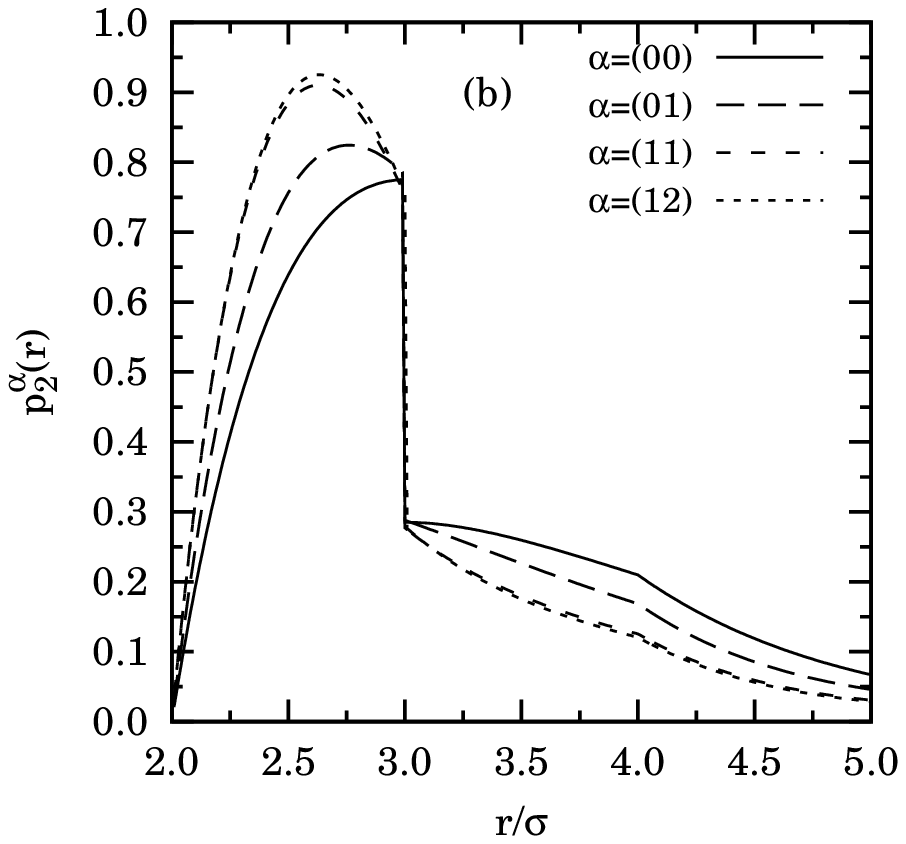}
\end{center}
\caption{(a) Plot of $p_1^{(00)}(r)$ (---), $p_1^{(01)}(r)$ (-- --), $p_1^{(11)}(r)$ (- - -), and $p_1^{(12)}(r)$ ($\cdots$) at $T^*=1$ and $p^*=1$. (b) Plot of $p_2^{(00)}(r)$ (---), $p_2^{(01)}(r)$ (-- --), $p_2^{(11)}(r)$ (- - -), and $p_2^{(12)}(r)$ ($\cdots$)  for the SW fluid ($\lambda=3$) at $T^*=1$ and $p^*=1$.}
\label{fig:pr}
\end{figure}

\begin{figure}
\begin{center}
\includegraphics[width=0.45\columnwidth]{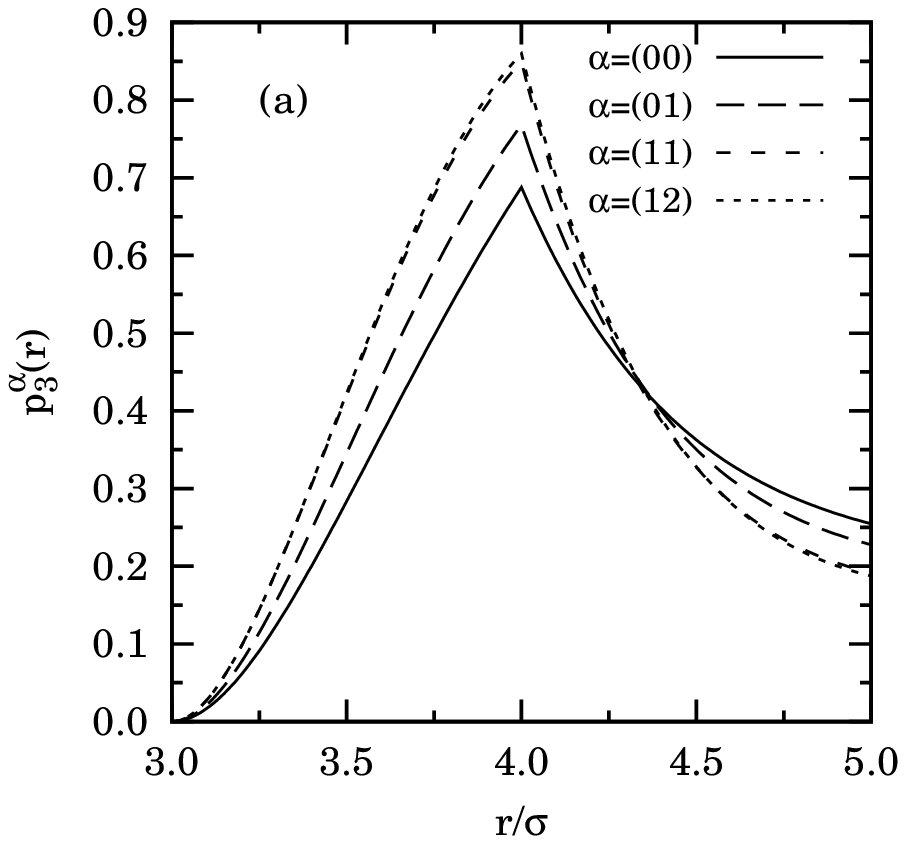}
\includegraphics[width=0.45\columnwidth]{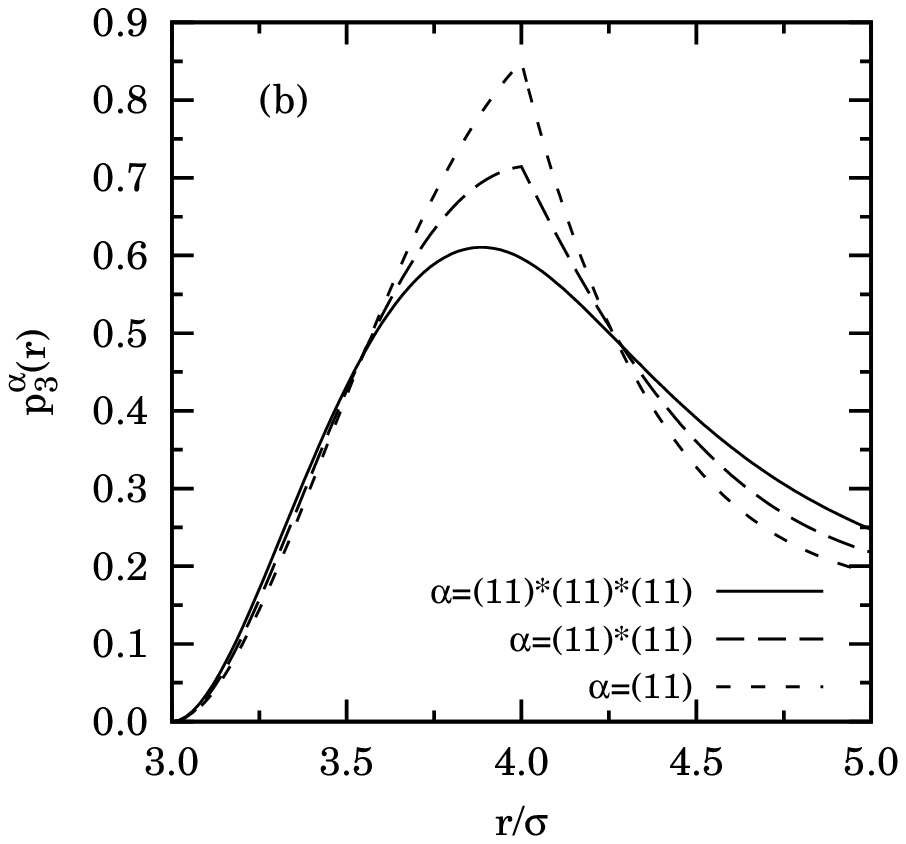}
\end{center}
\caption{(a) Plot of $p_3^{(00)}(r)$ (---), $p_3^{(01)}(r)$ (-- --), $p_3^{(11)}(r)$ (- - -), and $p_3^{(12)}(r)$ ($\cdots$) at $T^*=1$ and $p^*=1$. (b) Plot of $(p_1^{(11)}*p_1^{(11)}*p_1^{(11)})(r)$ (---), $(p_1^{(11)}*p_2^{(11)})(r)$ (-- --),  and $p_3^{(11)}(r)$ (- - -) for the SW fluid ($\lambda=3$) at $T^*=1$ and $p^*=1$.}
\label{fig:p3}
\end{figure}

Before comparing with MC simulations, let us analyze the convergence of the approximations presented in Sects.\ \ref{sec3.1}--\ref{sec3.3}.  Figure \ref{fig:pr}(a) shows $p_1^{(00)}(r)$, $p_1^{(01)}(r)$, $p_1^{(11)}(r)$, and $p_1^{(12)}(r)$ (the latter quantity not explicitly defined in  Sect.\ \ref{sec3.1}) at the state $T^*=1$, $p^*=1$.
We observe that the second-order
approximation $p_1^{(11)}(r)$ is  almost
indistinguishable from the third-order one $p_1^{(12)}(r)$. We have also checked that the fourth order approximation $p_1^{(22)}(r)$ differs from the third-order one by about $0.1\%$. Therefore, we can conclude that convergence has been practically reached already at second order.

The 2nd nn functions $p_2^{(00)}(r)$, $p_2^{(01)}(r)$, $p_2^{(11)}(r)$, and $p_2^{(12)}(r)$ at the same thermodynamic state are plotted in Fig.\ \ref{fig:pr}(b). Again we find that a good convergence has been reached with the second-order approximation $p_2^{(11)}(r)$.

Figure \ref{fig:p3}(a) is equivalent to Fig.\ \ref{fig:pr} but for the 3rd nn distribution. Once more, we observe that the second-order approximation $p_3^{(11)}(r)$ is hardly distinguishable from the third-order approximation $p_3^{(12)}(r)$. The convolution approximations
$(p_1^{(11)}*p_1^{(11)}*p_1^{(11)})(r)$ and $(p_1^{(11)}*p_2^{(11)})(r)$ are compared with $p_3^{(11)}(r)$ in Fig.\ \ref{fig:p3}(b).
As expected, the convolution functions $p_1*p_1*p_1$ and $p_1*p_2$ are only qualitatively correct in describing the 3rd nn distribution. In fact,  $p_1*p_1*p_1$ fails in capturing the kink of $p_3$ at $r=4$. All of this confirms that, in principle, Eq.\ \eqref{RDF2} is a better approximation than Eq.\ \eqref{RDF2b} but it is worse than any of Eqs.\ \eqref{RDF3abc}, at least in the range $1<r<4$.

Once we have seen that the second-order approximations for $p_1(r)$, $p_2(r)$, and $p_3(r)$ represent a good balance between simplicity and accuracy, we consider now the RDF  and compare the theoretical approximations of Sect.\ \ref{sec3.4} with our own canonical MC simulations (with $N=1024$ particles).

\begin{figure}
\begin{center}
\includegraphics[width=0.45\columnwidth]{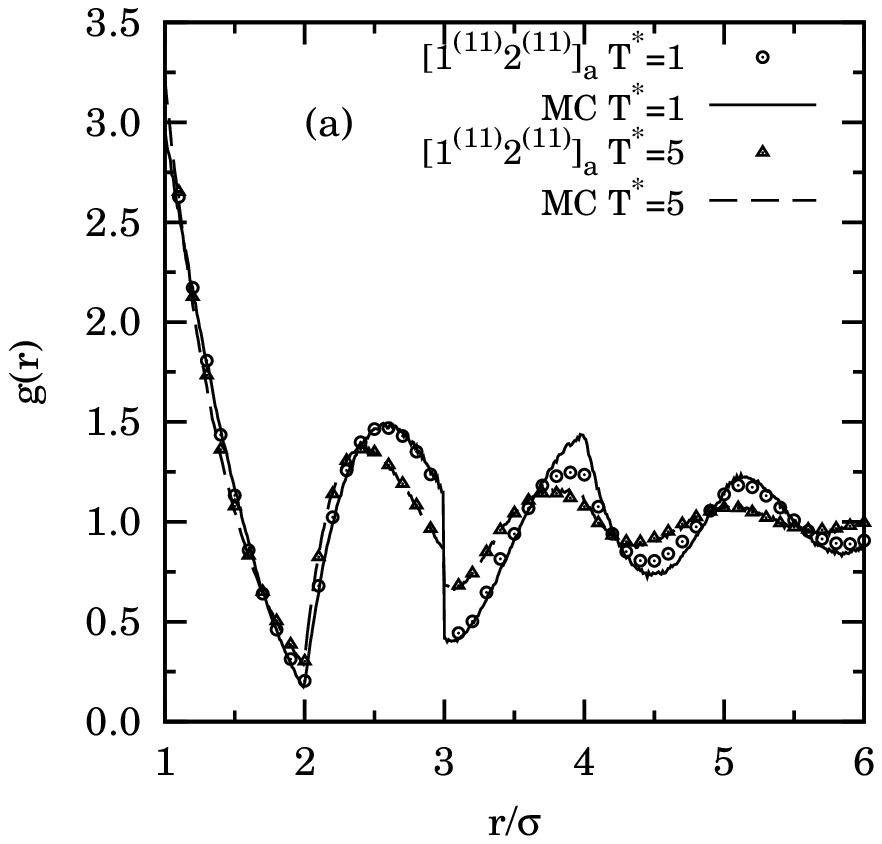}\includegraphics[width=0.45\columnwidth]{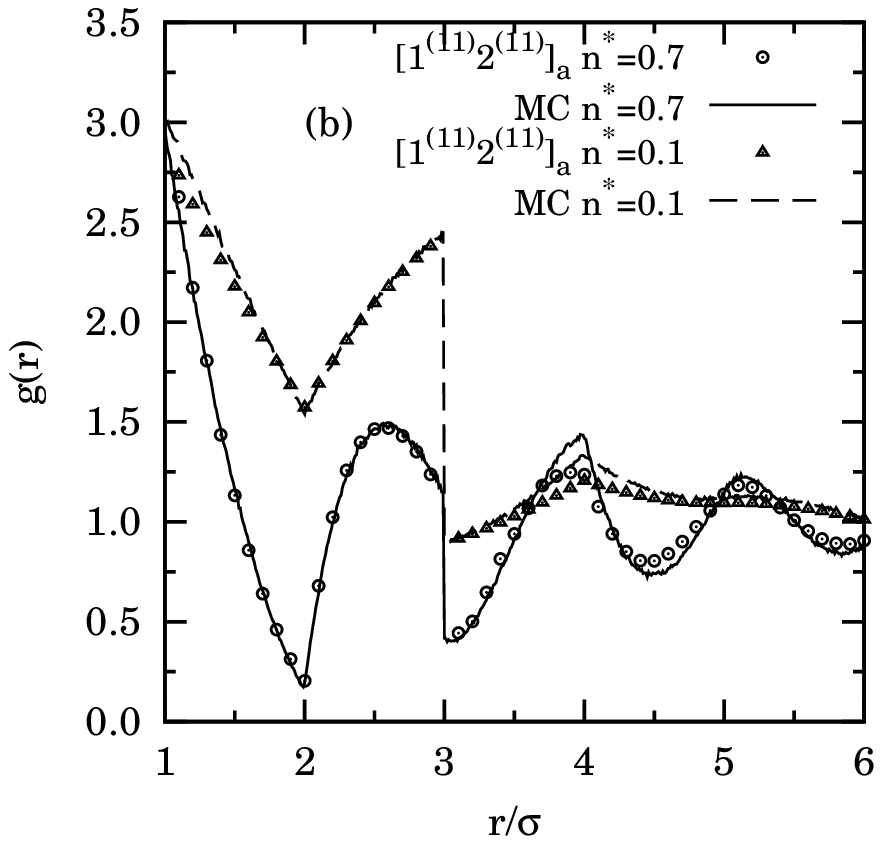}
\end{center}
\caption{Plot of $g(r)$ as obtained from MC simulations ({---} and -- --) and from the approximation $[1^{(11)}2^{(11)}]_a$ [see Eq.\ \eqref{RDF2}] ($\circ$ and $\vartriangle$) for the SW fluid ($\lambda=3$) at (a) $n^*=0.7$ and two temperatures ($T^*=1$ and $T^*=5$, respectively) and (b) $T^*=1$ and two densities ($n^*=0.7$ and $n^*=0.1$, respectively).}
\label{fig:grn&T}
\end{figure}

\begin{figure}
\begin{center}
\includegraphics[width=0.45\columnwidth]{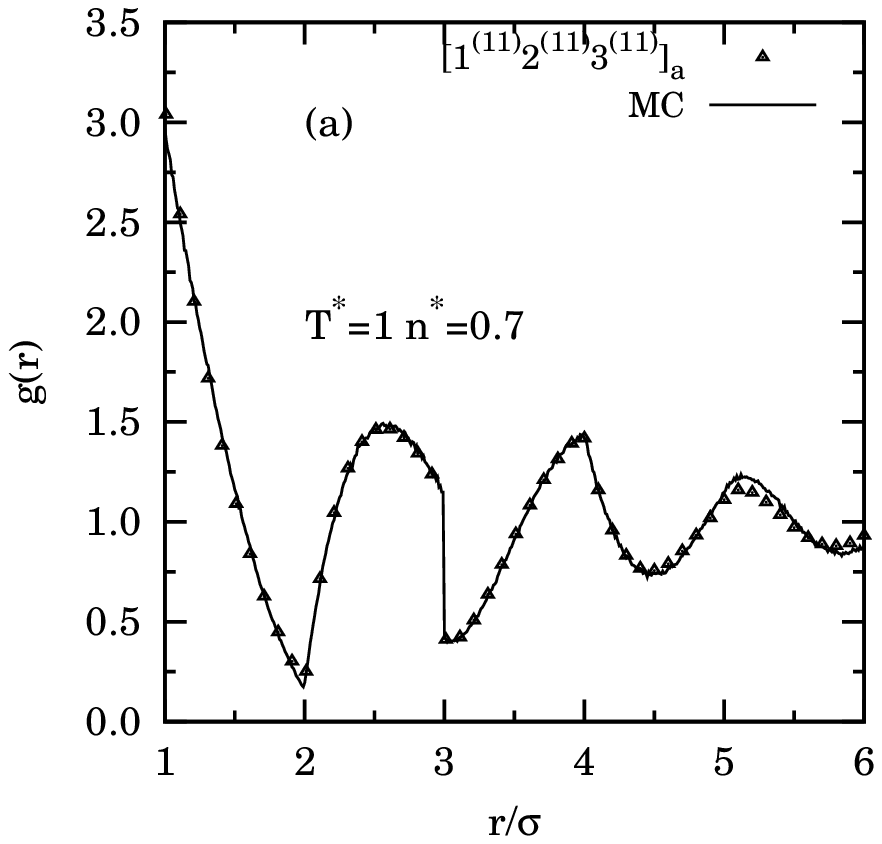}\includegraphics[width=0.45\columnwidth]{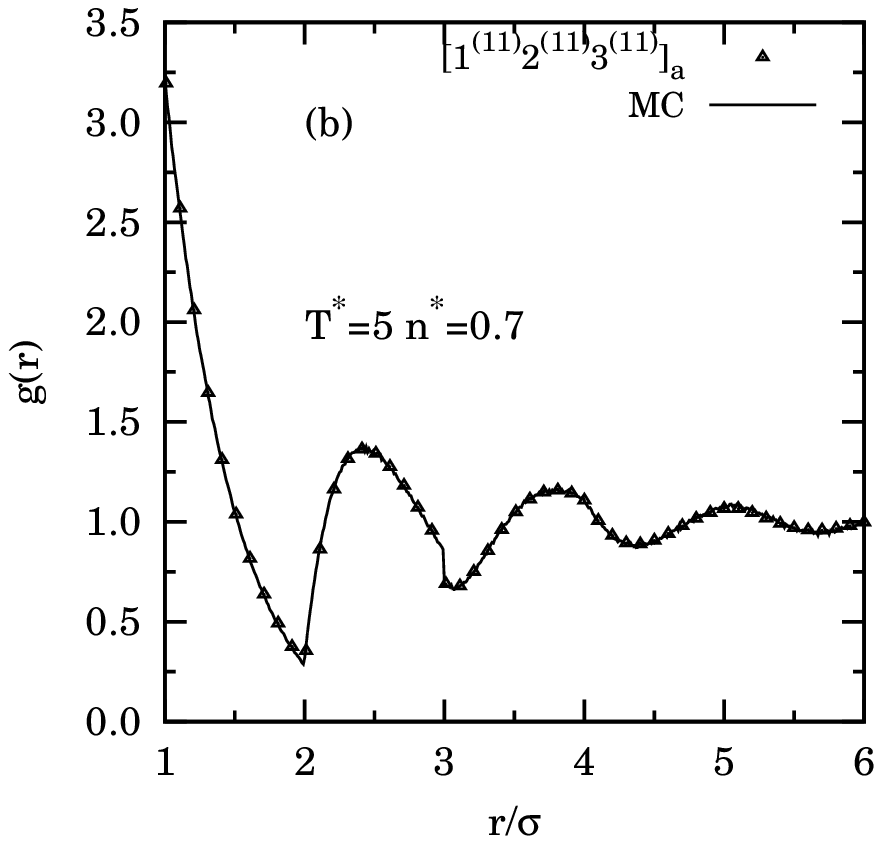}\\
\includegraphics[width=0.45\columnwidth]{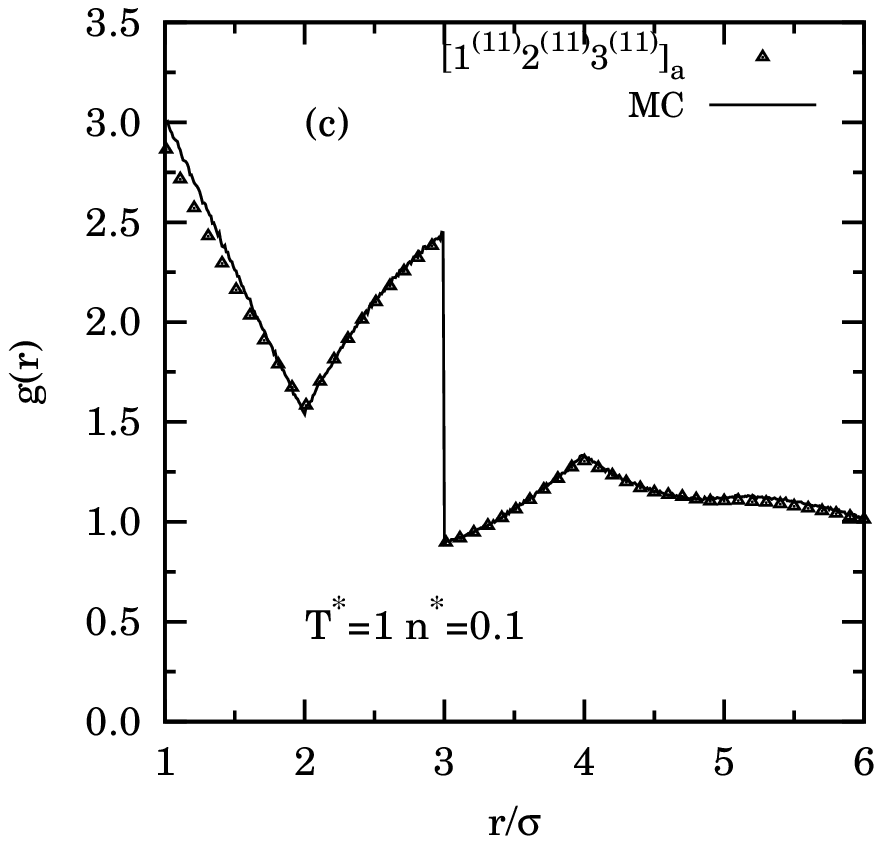}\includegraphics[width=0.45\columnwidth]{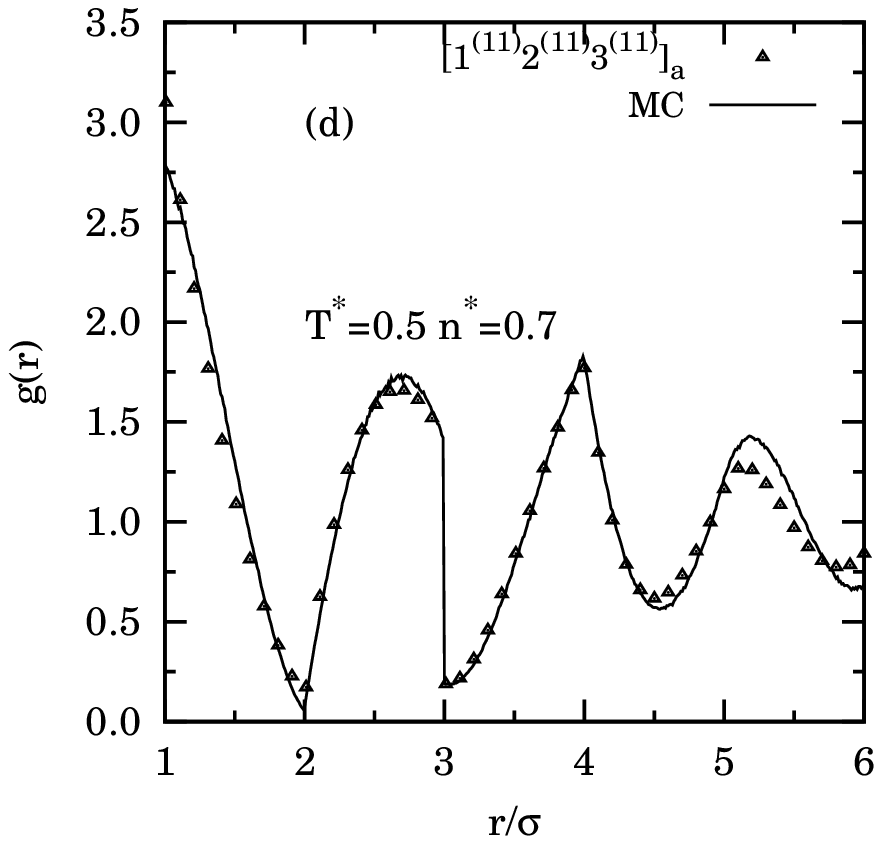}
\end{center}
\caption{Plot of $g(r)$ as obtained from MC simulations ({---}) and from the approximation $[1^{(11)}2^{(11)}3^{(11)}]_a$ [see Eq.\ \eqref{RDF3a}] ($\vartriangle$) for the SW fluid ($\lambda=3$) at (a) $T^*=1$ and $n^*=0.7$, (b) $T^*=5$ and $n^*=0.7$, (c) $T^*=1$ and $n^*=0.1$, and (d) $T^*=0.5$ and $n^*=0.7$.}
\label{fig:grn&T_p3}
\end{figure}

In Fig.\ \ref{fig:grn&T}  we show the RDF calculated from the approximation $[1^{(11)}2^{(11)}]_a$ [see Eq.\ \eqref{RDF2}] compared
with our MC simulations, at several values
of the reduced temperature and  density. It is apparent that the approximation $[1^{(11)}2^{(11)}]_a$ works very well in the region $1<r<3$, where $ng(r)=p_1(r)+p_2(r)$, and keeps being generally good for larger distances, even though $p_3$ is approximated by $p_1*p_2$.

In view of Fig.\ \ref{fig:p3}(b), the quality of the approximations in the region  $r>3$ is expected to improve if $p_3$ is approximated with independence of $p_1$ and $p_2$, as done in Eqs.\ \eqref{RDF3abc}. This is confirmed by Fig.\ \ref{fig:grn&T_p3}, where the approximation $[1^{(11)}2^{(11)}3^{(11)}]_a$ [see Eq.\ \eqref{RDF3a}] is compared with MC data for the same three states as in Fig.\ \ref{fig:grn&T} plus the more stringent state $T^*=0.5$ and $n^*=0.7$.
A very good agreement is observed, although, not surprisingly, the quality of the approximation worsens as the temperature decreases and the density increases [see Fig.\ \ref{fig:grn&T_p3}(d)].

\subsection{Thermodynamic Properties}
\label{sec4.2}

The approximations for the 1st, 2nd, and 3rd nn probability distribution functions and for the RDF worked out in Sect.\ \ref{sec:approx} can also be used to obtain the thermodynamic properties, as presented in Sect.\ \ref{subsubsec:thermo}. We focus on the equation of state (i.e., the relationship between pressure, density, and temperature), the isothermal susceptibility, and the excess internal energy per particle. Given the approximate character of our proposals, the results will in general depend on the route followed to obtain those thermodynamic quantities. In fact, the degree of thermodynamic inconsistency will be used as a test of our approach.

The most direct way of determining the equation of state is as $n=\ell/\widetilde{p}_\ell$, with $\ell=1,2,3$, in accordance with Eqs.\ \eqref{RDF2ab} and \eqref{RDF3abc}. This gives the number density as an explicit function of pressure  and temperature, i.e., $n(\beta,\beta p)$. If we prefer to express the pressure as a function of density and temperature, $p(n,T)$, we need to solve numerically the equation $n=n(\beta,\beta p)$. In either choice of independent variables, the compressibility factor is obtained as $Z\equiv \beta p/n$. We will use the superscript (A), i.e., $n^{(\mathrm{A})}$ and  $Z^{(\mathrm{A})}$, to denote this ``direct'' route to the equation of state. From it, the excess internal energy per particle ($u$) and the isothermal susceptibility ($\chi_T$) can be obtained via the thermodynamic properties \eqref{ct3} and \eqref{chia}, respectively, namely
\begin{subequations}
\bq
\label{4.3}
u^{(\mathrm{A})}(\beta,\beta p)=\beta p\int_0^1 \rmd t\,\left(\frac{\partial 1/n^{(\mathrm{A})}(\beta, t\beta p)}{\partial \beta}\right)_{\beta p},
\eq
\bq
\label{4.4}
{\chi_T^{(\mathrm{A})}}=\left(\frac{\partial n^{(\mathrm{A})}}{\partial\beta p}\right)_\beta=\frac{1}{\left({\partial n Z^{(\mathrm{A})}}/{\partial n}\right)_T}.
\eq
\end{subequations}

Alternatively,  $u$, $Z$, and $\chi_T$ can be obtained from the  energy (e), virial (v), and compressibility (c) routes, respectively, as given by Eqs.\ \eqref{er}, \eqref{vr}, and \eqref{chib}. In particular, for the SW potential \eqref{SWpotential} one has
\bq
\label{4.5}
\frac{u^{(\mathrm{e})}}{\epsilon}=-\int_1^\lambda \rmd r\, p_1(r)-\int_2^\lambda \rmd r\, p_2(r),
\eq
\bq
\label{4.6}
Z^{(\mathrm{v})}&=&1+n\left[g(1^+)-\lambda\left(\rme^{\beta\epsilon}-1\right)g(\lambda^+)\right]\nn
&=&1+p_1(1^+)-\lambda\left(\rme^{\beta\epsilon}-1\right)\left[p_1(\lambda^+)+p_2(\lambda^+)\right].
\eq
As for the isothermal susceptibility, application of the approximations \eqref{RDF2ab} and \eqref{RDF3abc} into Eq.\ \eqref{chib} yields
\begin{subequations}
\label{4.7abc}
\bq
\label{4.7a}
\chi_T^{(\mathrm{c})}=\frac{3\, \widetilde{\!\widetilde{p}}_3}{\widetilde{p}_3^2}
-\frac{2 (\widetilde{p}_1+\widetilde{p}_2)}{\widetilde{p}_3}-1,
\eq
\bq
\label{4.7b}
\chi_T^{(\mathrm{c})}=\frac{2\, \widetilde{\!\widetilde{p}}_2}{\widetilde{p}_2^2}
-\frac{2\widetilde{p}_1}{\widetilde{p}_2}-1,
\eq
\bq
\label{4.7c}
\chi_T^{(\mathrm{c})}=\frac{\,\widetilde{\!\widetilde{p}}_1}{\widetilde{p}_1^2}
-1,
\eq
\end{subequations}
where
\bq
\widetilde{\!\widetilde{p}}_\ell\equiv \left.\frac{\partial^2 \widehat{p}_\ell(s)}{\partial s^2}\right|_{s=0}=\int_0^\infty \rmd r\, r^2p_\ell (r).
\eq
Equation \eqref{4.7a} applies to Eq.\ \eqref{RDF3a}, while Eq.\ \eqref{4.7b} applies to Eqs.\ \eqref{RDF2} and \eqref{RDF3b}, and Eq.\ \eqref{4.7c} applies to Eqs.\ \eqref{RDF2b} and \eqref{RDF3c}.
{From} Eq.\ \eqref{chia} and $\chi_T^{(\mathrm{c})}$, the compressibility factor can be obtained as
\bq
\label{4.8}
Z^{(\mathrm{c})}(\beta,\beta p)=\frac{1}{\int_0^1\rmd t\, \chi_T^{(\mathrm{c})}(\beta,t \beta p)}.
\eq

The thermodynamic quantities $Z^{(\mathrm{A})}$, $u^{(\mathrm{A})}$, $\chi_T^{(\mathrm{A})}$, $Z^{(\mathrm{v})}$, $u^{(\mathrm{e})}$, $\chi_T^{(\mathrm{c})}$, and $Z^{(\mathrm{c})}$ are common to those approximations \eqref{RDF2ab} and \eqref{RDF3abc} having the same denominator of the form $1-\widehat{p}_\ell(s)$.
In our notation, this means that $[1^{(\alpha_1)}2^{(\alpha_2)}]_a=[1^{(\alpha_1)}2^{(\alpha_2)}3^{(\alpha_2)}]_b$ and $[1^{(\alpha_1)}2^{(\alpha_2)}]_b=[1^{(\alpha_1)}2^{(\alpha_2)}3^{(\alpha_3)}]_c$ in what concerns the thermodynamic properties. Thus, here we will refer to the thermodynamic properties associated with the three approximations $[1^{(11)}2^{(11)}3^{(11)}]_{a,b,c}$.

\begin{figure}
\begin{center}
\includegraphics[width=0.45\columnwidth]{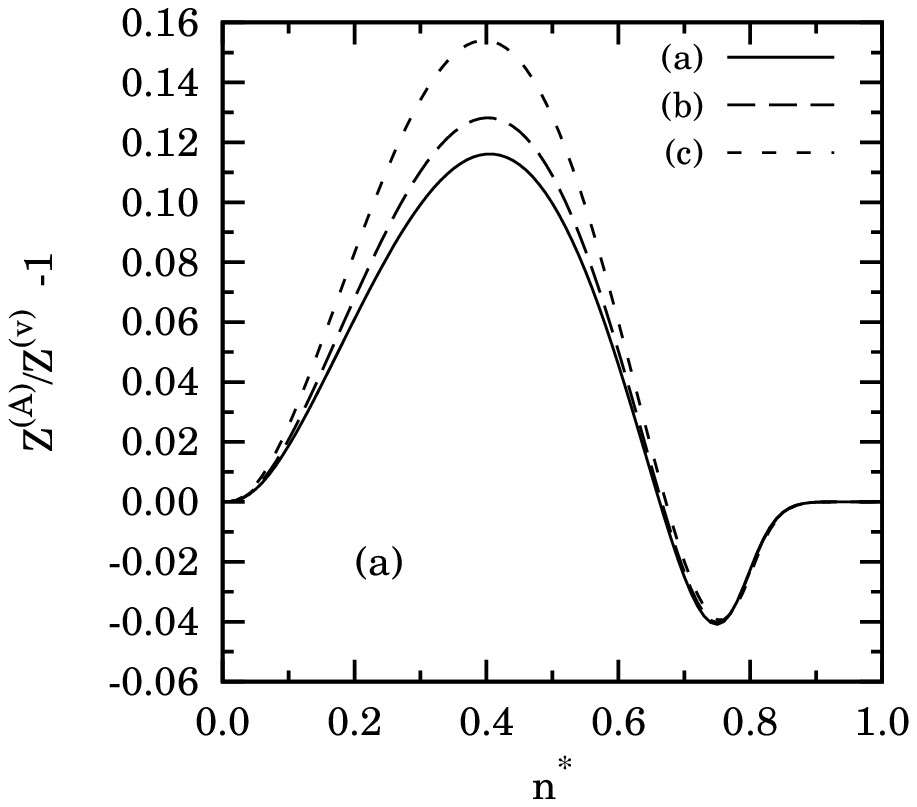}\includegraphics[width=0.45\columnwidth]{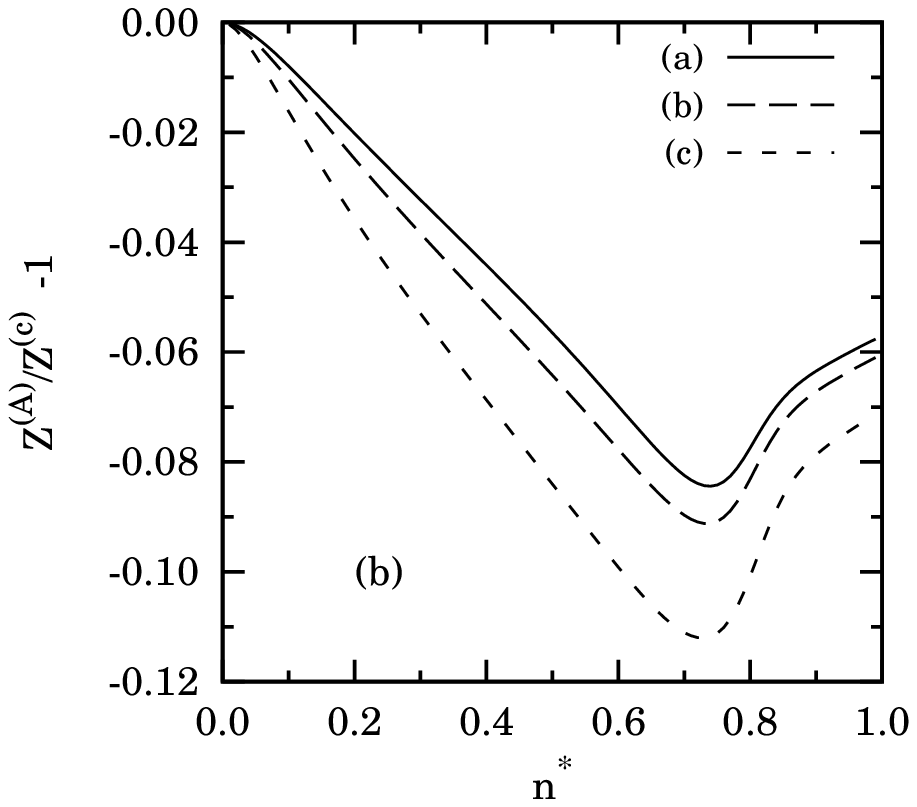}\\
\includegraphics[width=0.45\columnwidth]{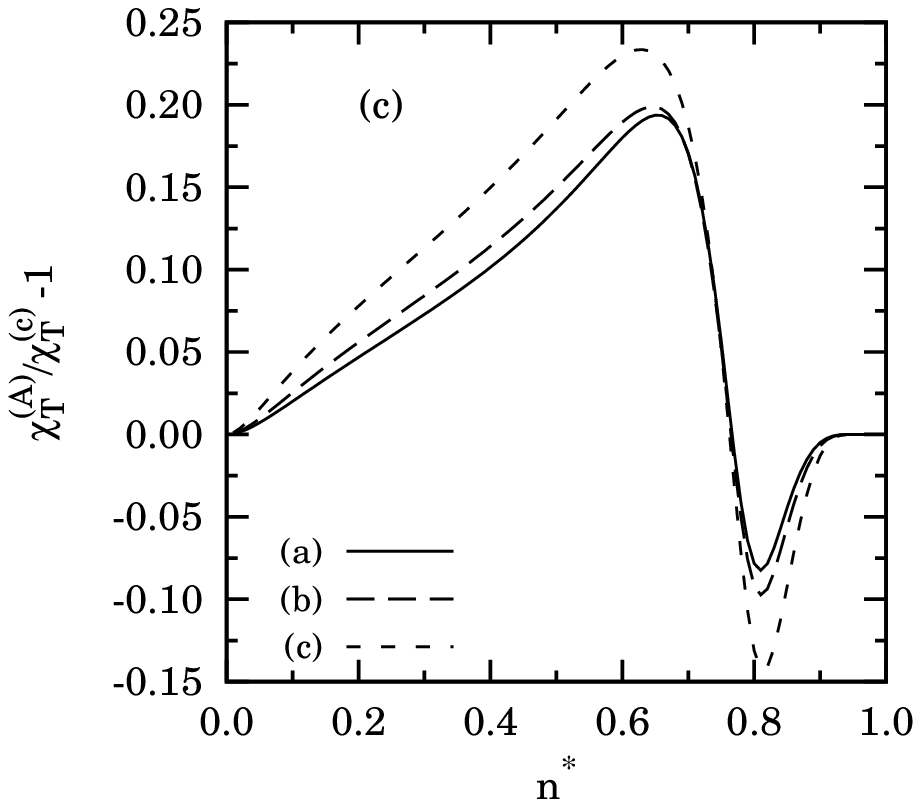}\includegraphics[width=0.45\columnwidth]{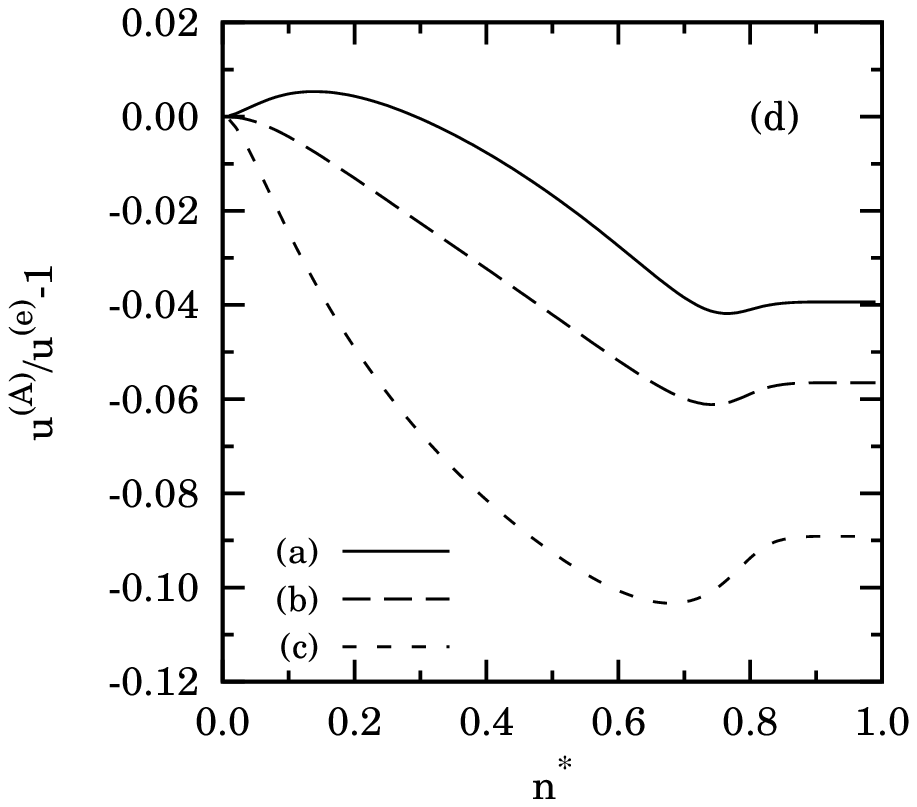}
\end{center}
\caption{Plot of the relative differences (a) $Z^{(\mathrm{A})}/Z^{(\mathrm{v})}-1$, (b) $Z^{(\mathrm{A})}/Z^{(\mathrm{c})}-1$, (c) $\chi_T^{(\mathrm{A})}/\chi_T^{(\mathrm{c})}-1$, and (d) $u^{(\mathrm{A})}/u^{(\mathrm{e})}-1$ as functions of density for the SW fluid ($\lambda=3$) at $T^*=1$. The curves correspond to the approximations $[1^{(11)}2^{(11)}3^{(11)}]_{a}$ (---), $[1^{(11)}2^{(11)}3^{(11)}]_{b}$ (-- --), and $[1^{(11)}2^{(11)}3^{(11)}]_{c}$ (- - -).}
\label{fig:SWthermod}
\end{figure}

Figure \ref{fig:SWthermod} presents thermodynamic consistency tests for the different routes within the approximations $[1^{(11)}2^{(11)}3^{(11)}]_{a,b,c}$. As expected, $[1^{(11)}2^{(11)}3^{(11)}]_{a}$ is the most consistent approximation, the thermodynamic quantities deviating typically less than $10\%$ at the relatively low temperature $T^*=1$.

\begin{table}
\caption{Some thermodynamic quantities for the SW fluid ($\lambda=3$) at several values of $T^*$ and  $n^*$.}
\label{tab:tq}
\begin{center}
\begin{tabular}{llllllll}
\noalign{\smallskip}\hline
Method & $Z^{(\mathrm{A})}$ & $Z^{(\mathrm{v})}$ & $Z^{(\mathrm{c})}$& $\chi_T^{(\mathrm{A})}$ & $\chi_T^{(\mathrm{c})}$ &$u^{(\mathrm{A})}/\epsilon$&
$u^{(\mathrm{e})}/\epsilon$\\
\noalign{\smallskip}\hline\noalign{\smallskip}
\multicolumn{8}{c}{$T^*=1, n^*=0.1$}\\
\noalign{\smallskip}\hline\noalign{\smallskip}
MC& \multicolumn{3}{c}{$0.8402(6)$} & \multicolumn{2}{c}{---}  &\multicolumn{2}{c}{$-0.434(1)$}\\
$[1^{(11)}2^{(11)}3^{(11)}]_a$&$0.8427$&$0.8271$&$0.8494$&$1.3381$&$1.3117$&$-0.4273$&$-0.4252$\\
$[1^{(11)}2^{(11)}3^{(11)}]_b$&$0.8441$&$0.8269$&$0.8529$&$1.3309$&$1.2978$&$-0.4241$&$-0.4259$\\
$[1^{(11)}2^{(11)}3^{(11)}]_c$&$0.8474$&$0.8263$&$0.8614$&$1.3151$&$1.2667$&$-0.4169$&$-0.4274$\\
\noalign{\smallskip}\hline\noalign{\smallskip}
\multicolumn{8}{c}{$T^*=1, n^*=0.4$}\\
\noalign{\smallskip}\hline\noalign{\smallskip}
MC& \multicolumn{3}{c}{$0.785(1)$} & \multicolumn{2}{c}{---}  &\multicolumn{2}{c}{$-1.168(1)$}\\
$[1^{(11)}2^{(11)}3^{(11)}]_a$&$0.8087$&$0.7246$&$0.8461$&$0.9596$&$0.8713$&$-1.1213$&$-1.1299$\\
$[1^{(11)}2^{(11)}3^{(11)}]_b$&$0.8200$&$0.7269$&$0.8643$&$0.9347$&$0.8389$&$-1.1014$&$-1.1381$\\
$[1^{(11)}2^{(11)}3^{(11)}]_c$&$0.8450$&$0.7322$&$0.9073$&$0.8822$&$0.7672$&$-1.0618$&$-1.1559$\\
\noalign{\smallskip}\hline\noalign{\smallskip}
\multicolumn{8}{c}{$T^*=1, n^*=0.7$}\\
\noalign{\smallskip}\hline\noalign{\smallskip}
MC&\multicolumn{3}{c}{$1.575(3)$} & \multicolumn{2}{c}{---}&\multicolumn{2}{c}{$-1.752(1)$}\\
$[1^{(11)}2^{(11)}3^{(11)}]_a$&$1.6263$&$1.6680$&$1.7725$&$0.1268$&$0.1083$&$-1.6772$&$-1.7442$\\
$[1^{(11)}2^{(11)}3^{(11)}]_b$&$1.6480$&$1.6875$&$1.8103$&$0.1278$&$0.1092$&$-1.6440$&$-1.7487$\\
$[1^{(11)}2^{(11)}3^{(11)}]_c$&$1.6980$&$1.7324$&$1.9102$&$0.1304$&$0.1098$&$-1.5774$&$-1.7587$\\
\noalign{\smallskip}\hline\noalign{\smallskip}
\multicolumn{8}{c}{$T^*=5, n^*=0.1$}\\
\noalign{\smallskip}\hline\noalign{\smallskip}
MC& \multicolumn{3}{c}{$1.0669(1)$} & \multicolumn{2}{c}{---}  &\multicolumn{2}{c}{$-0.2373(4)$}\\
$[1^{(11)}2^{(11)}3^{(11)}]_a$&$1.0671$&$1.0670$&$1.0674$&$0.8724$&$ 0.8715$&$-0.2368$&$-0.2368$\\
$[1^{(11)}2^{(11)}3^{(11)}]_b$&$1.0671$&$1.0670$&$1.0676$&$0.8723$&$0.8711$&$-0.2367$&$-0.2368$\\
$[1^{(11)}2^{(11)}3^{(11)}]_c$&$1.0671$&$1.0670$&$1.0681$&$0.8723$&$0.8698$&$-0.2366$&$-0.2368$\\
\noalign{\smallskip}\hline\noalign{\smallskip}
\multicolumn{8}{c}{$T^*=5, n^*=0.4$}\\
\noalign{\smallskip}\hline\noalign{\smallskip}
MC& \multicolumn{3}{c}{$1.4733(5)$} & \multicolumn{2}{c}{---}  &\multicolumn{2}{c}{$-0.9230(8)$}\\
$[1^{(11)}2^{(11)}3^{(11)}]_a$&$1.4769$&$1.4738$&$1.4840$&$0.4230$&$0.4172$&$-0.9188$&$-0.9217$\\
$[1^{(11)}2^{(11)}3^{(11)}]_b$&$1.4775$&$1.4740$&$1.4875$&$0.4226$&$0.4147$&$-0.9169$&$-0.9220$\\
$[1^{(11)}2^{(11)}3^{(11)}]_c$&$1.4789$&$1.4746$&$1.4973$&$0.4216$&$0.4075$&$-0.9116$&$-0.9225$\\
\noalign{\smallskip}\hline\noalign{\smallskip}
\multicolumn{8}{c}{$T^*=5, n^*=0.7$}\\
\noalign{\smallskip}\hline\noalign{\smallskip}
MC& \multicolumn{3}{c}{$2.940(2)$} & \multicolumn{2}{c}{---} & \multicolumn{2}{c}{$-1.683(1)$}\\
$[1^{(11)}2^{(11)}3^{(11)}]_a$&$2.9592$&$2.9613$&$2.9915$&$0.09464$&$0.09345$&$-1.6697$&$-1.6843$\\
$[1^{(11)}2^{(11)}3^{(11)}]_b$&$2.9602$&$2.9620$&$3.0040$&$0.09470$&$0.09317$&$-1.6623$&$-1.6845$\\
$[1^{(11)}2^{(11)}3^{(11)}]_c$&$2.9627$&$2.9638$&$3.0397$&$0.09486$&$0.09237$&$-1.6423$&$-1.6849$\\
\noalign{\smallskip}\hline
\end{tabular}
\end{center}

\end{table}

\begin{figure}
\begin{center}
\includegraphics[width=0.45\columnwidth]{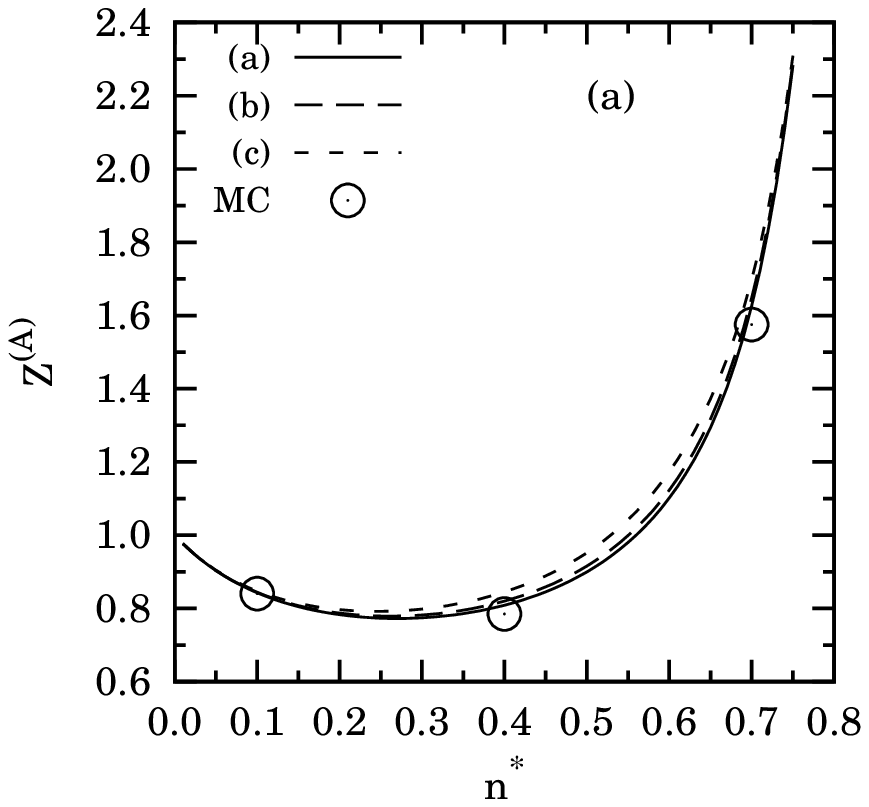}
\includegraphics[width=0.45\columnwidth]{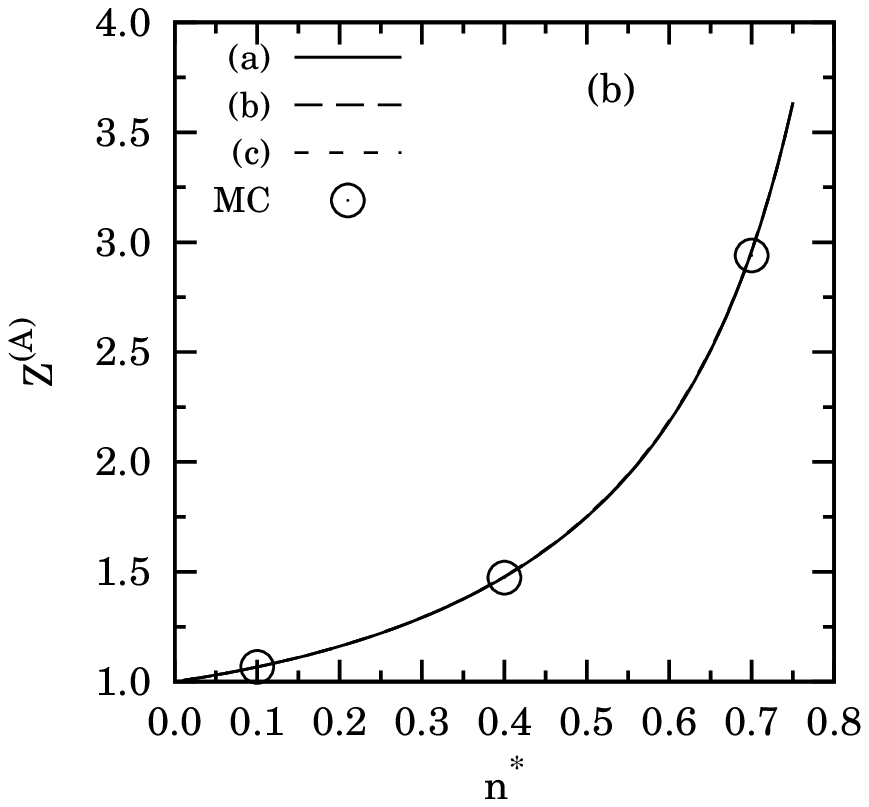}\\
\includegraphics[width=0.45\columnwidth]{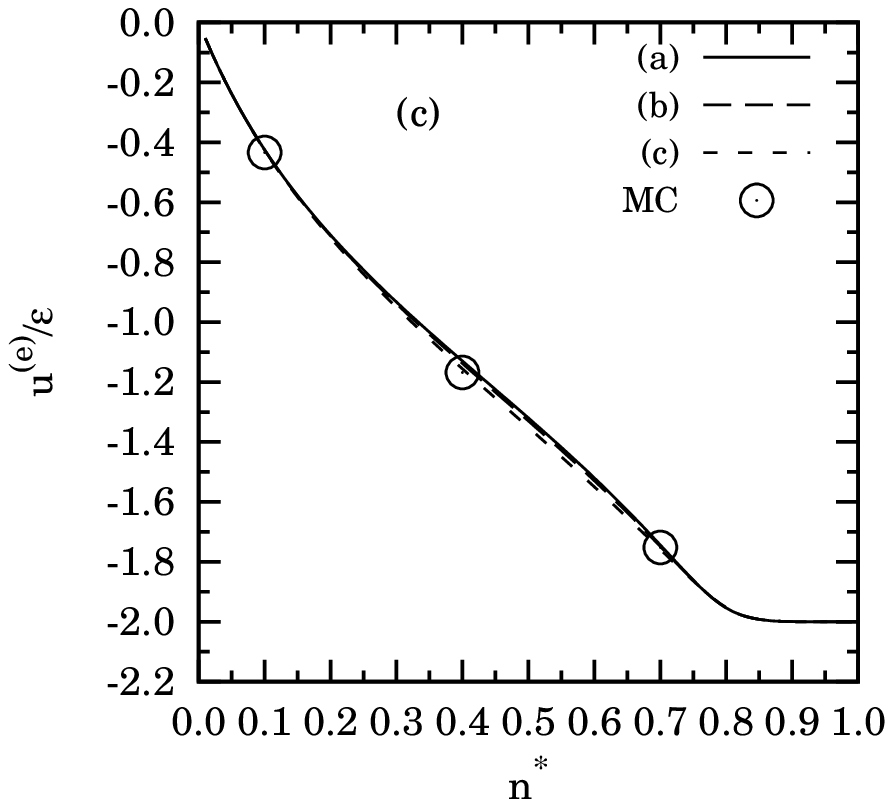}
\includegraphics[width=0.45\columnwidth]{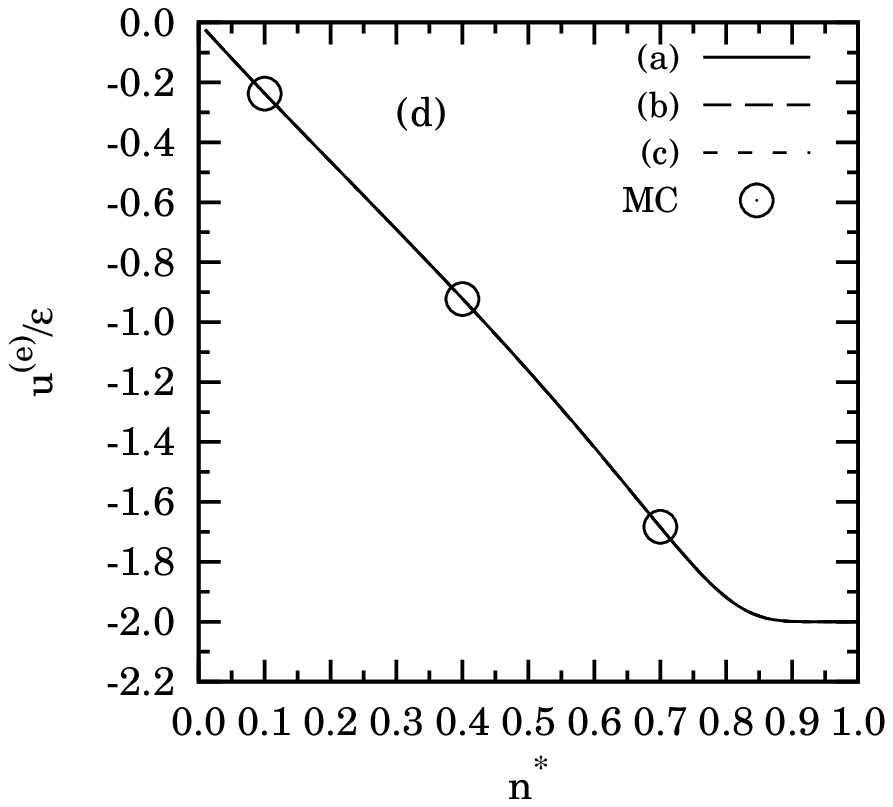}
\end{center}
\caption{Plot of the compressibility factor $Z$ [panels (a) and (b)] and of the
excess internal energy per particle $u$ [panels (c) and (d)] as functions of
density for the SW fluid ($\lambda=3$) at $T^*=1$ [panels (a) and (c)] and $T^*=5$ [panels (b) and (d)]. The curves are theoretical results $Z^{(\mathrm{A})}$ and $u^{(\mathrm{e})}$, as obtained from the approximations $[1^{(11)}2^{(11)}3^{(11)}]_{a}$ (---), $[1^{(11)}2^{(11)}3^{(11)}]_{b}$ (-- --), and $[1^{(11)}2^{(11)}3^{(11)}]_{c}$ (- - -), while the circles represent MC data. Note that in panels (b)--(d) the three approximations $[1^{(11)}2^{(11)}3^{(11)}]_{a,b,c}$ yield practically indistinguishable results.}
\label{fig:SWthermod1}
\end{figure}

The theoretical values are compared with MC simulation results for a few thermodynamic states in Table \ref{tab:tq}. We can observe that the best agreement in the case of the compressibility factor is generally reached with the direct route, $Z^{(\mathrm{A})}$, in the $[1^{(11)}2^{(11)}3^{(11)}]_{a}$ approximation. The compressibility route, however, tends to overestimate the value of $Z$. In what refers to the excess internal energy, the energy route is generally better than the direct route. By a fortuitous cancelation of errors, the approximations $[1^{(11)}2^{(11)}3^{(11)}]_{b,c}$ can in some cases outperform the approximation $[1^{(11)}2^{(11)}3^{(11)}]_{a}$ in estimating $u$.

In  Fig.\ \ref{fig:SWthermod1} we show the behavior of the
compressibility factor and of the excess internal energy per particle
as functions of density at temperatures $T^*=1$ and $T^*=5$. At the latter temperature the three approximations $[1^{(11)}2^{(11)}3^{(11)}]_{a,b,c}$ provide practically indistinguishable results. As can be observed, the agreement with our MC results is very satisfactory.

\section{The Two-Step Potential}
\label{sec:TS}
Let us now consider the following TS fluid defined by the potential
\bq \label{TSpotential}
\phi(r)=\left\{\begin{array}{ll}
\infty, & r<\sigma,\\
-\epsilon, & \sigma\leq r<\lambda_1\sigma,\\
-\epsilon_2, & \lambda_1\sigma\leq r<\lambda\sigma,\\
0, & \lambda\sigma\leq r.
\end{array}\right.
\eq
Clearly, for $\epsilon_2=\epsilon$, or $\lambda_1=1$, or $\lambda_1=\lambda$ we recover the SW fluid of Sect.\
\ref{sec:SW}.  More in general, playing with the
signs and the magnitudes of the two energy scales, $\epsilon$ and $\epsilon_2$, several classes of piece-wise constant potentials can be described \cite{SYHBO13,SYH12}. Here, we will restrict ourselves  just to the
case of full attraction with $\epsilon>\epsilon_2>0$. Analogously to the SW case, we define the reduced density and temperature as $n^*\equiv n\sigma$ and $T^*\equiv k_BT/\epsilon$, respectively.

Obviously, if $\lambda\leq 2$ the TS interaction cannot extend beyond 1st nn and therefore the exact solution described in Sect.\ \ref{sec:1nn} applies. On the other hand, if $2<\lambda\leq 3$, the interaction involves both 1st and 2nd nn, so that only approximate treatments are possible.
The aim of this section is to test the performance of the approximations of  Sect.\ \ref{sec:approx} against our MC simulations (again with $N=1024$ particles) in the case of the TS potential. To that end, we will fix $\lambda_1=1.5$, $\lambda=3$, and $\epsilon_2=\epsilon/2$. This means that the strength of the 2nd nn interactions is weaker in this TS potential than in the SW potential considered in Sect.\ \ref{sec:SW}. Therefore, at common values of $T^*$ and $n^*$, our approximations may be expected to be more accurate for the TS fluid than for the SW fluid.

\subsection{Structural Properties}
As happened in the case of the SW potential, the approximations described in Sects.\ \ref{sec3.1}--\ref{sec3.3} lend themselves to analytical implementations in the case of the TS potential. Moreover, due to the hard core at $r=\sigma$, Eq.\ \eqref{4.2} still applies.

\begin{figure}
\begin{center}
\includegraphics[width=0.45\columnwidth]{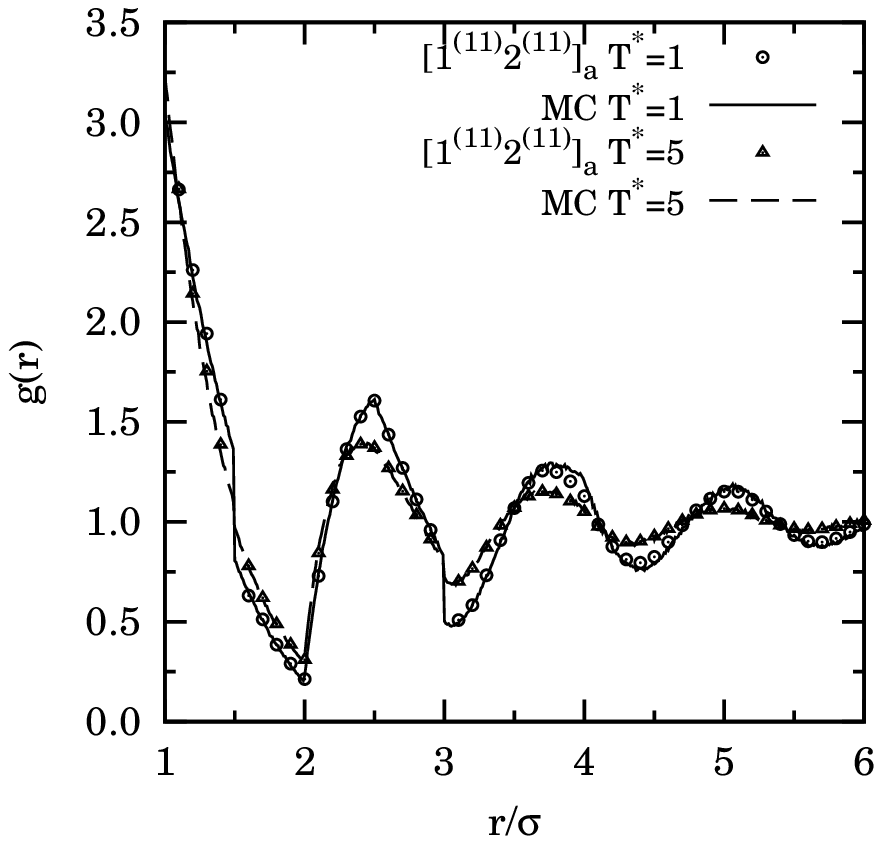}\includegraphics[width=0.45\columnwidth]{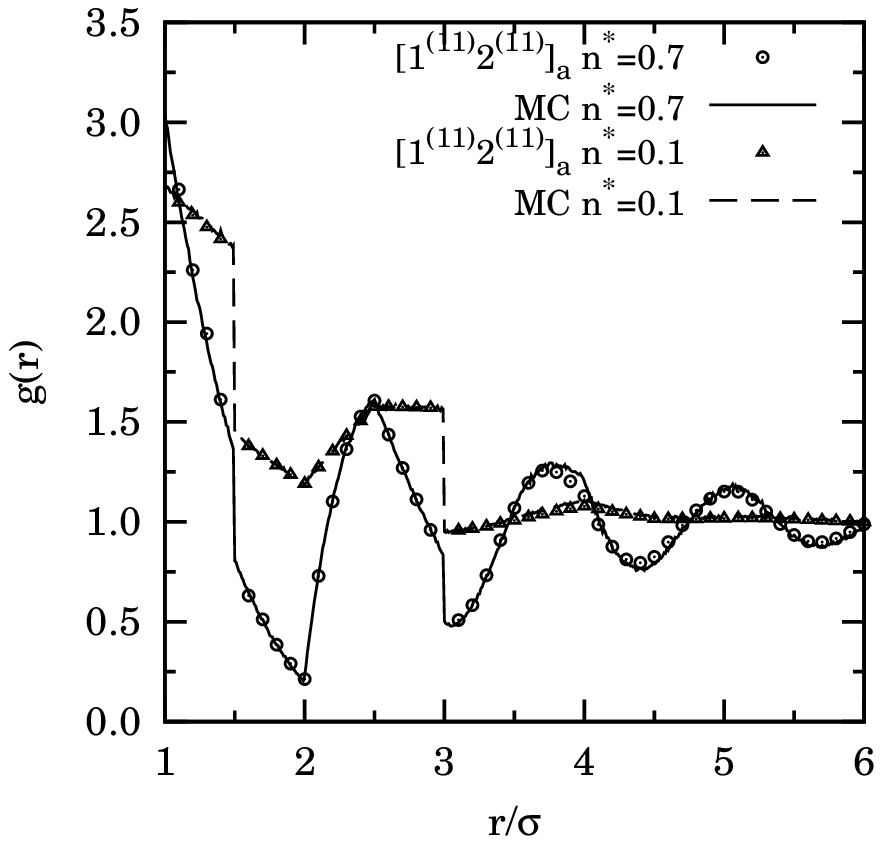}
\end{center}
\caption{Plot of $g(r)$ as obtained from MC simulations ({---} and -- --) and from the approximation $[1^{(11)}2^{(11)}]_a$ [see Eq.\ \eqref{RDF2}] ($\circ$ and $\vartriangle$) for the TS fluid ($2\lambda_1=\lambda=3$, $\epsilon_2/\epsilon=\frac{1}{2}$) at (a) $n^*=0.7$ and two temperatures ($T^*=1$ and $T^*=5$, respectively) and (b) $T^*=1$ and two densities ($n^*=0.7$ and $n^*=0.1$, respectively).}
\label{fig:grn&T(TS)}
\end{figure}

\begin{figure}\sidecaption
\resizebox{0.45\columnwidth}{!}{\includegraphics*{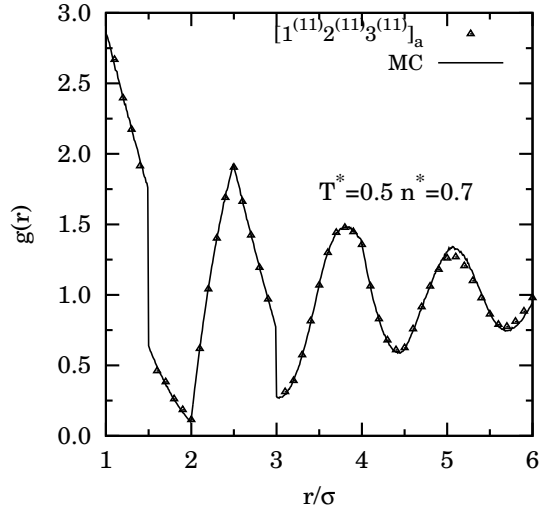}}
\caption{Plot of $g(r)$ as obtained from MC simulations ({---}) and from the approximation $[1^{(11)}2^{(11)}3^{(11)}]_a$ [see Eq.\ \eqref{RDF3a}] ($\vartriangle$) for the TS fluid ($2\lambda_1=\lambda=3$, $\epsilon_2/\epsilon=\frac{1}{2}$) at $T^*=0.5$ and $n^*=0.7$.}
\label{fig:TSgr(11)(11)(11)lt}
\end{figure}

Figure \ref{fig:grn&T(TS)} shows the  RDF
obtained from our approximation $[1^{(11)}2^{(11)}]_a$ at several values of $n^*$ and $T^*$  and tests it with the
result of our MC simulations. From comparison with Fig.\ \ref{fig:grn&T}, we see that, as expected, the agreement between the
theoretical approximation $[1^{(11)}2^{(11)}]_a$ and the MC simulations
improves with respect to the SW case treated in Sect.\ \ref{sec:SW}. The more sophisticated approximation $[1^{(11)}2^{(11)}3^{(11)}]_a$ does an even better job (not shown).
The improvement of the theoretical approximation $[1^{(11)}2^{(11)}3^{(11)}]_a$ when applied to the TS fluid rather than to the SW fluid is confirmed by Fig.\ \ref{fig:TSgr(11)(11)(11)lt}  at a low temperature ($T^*=0.5$) and high density ($n^*=0.7$) [compare with Fig.\ \ref{fig:grn&T_p3}(d)].

\begin{figure}
\begin{center}
\includegraphics[width=0.45\columnwidth]{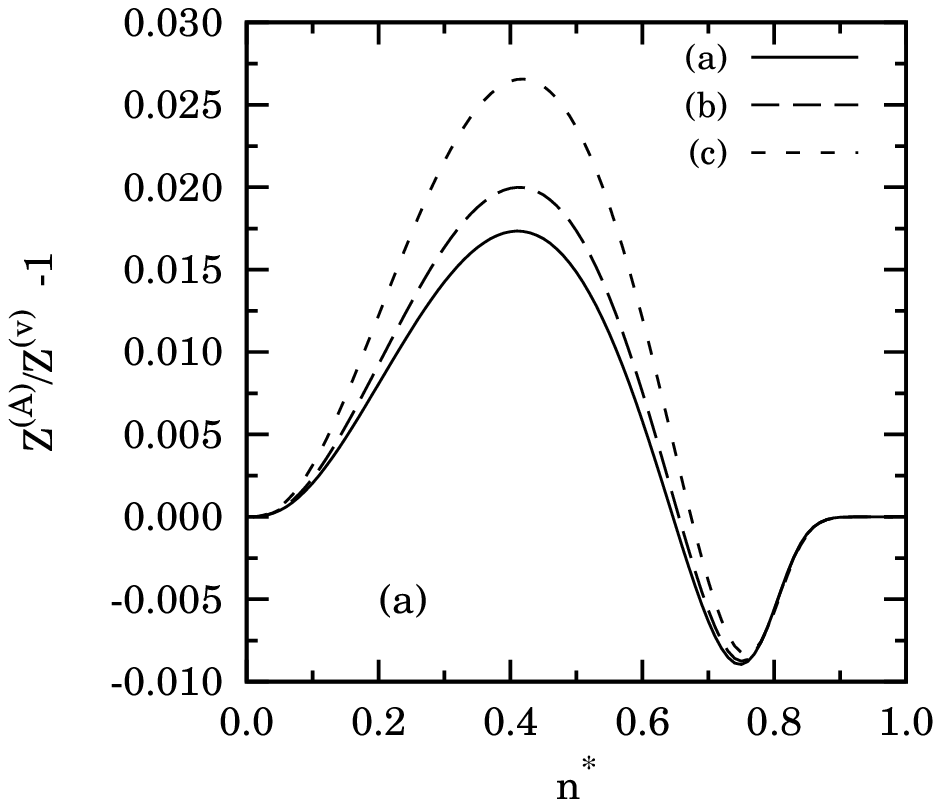}\includegraphics[width=0.45\columnwidth]{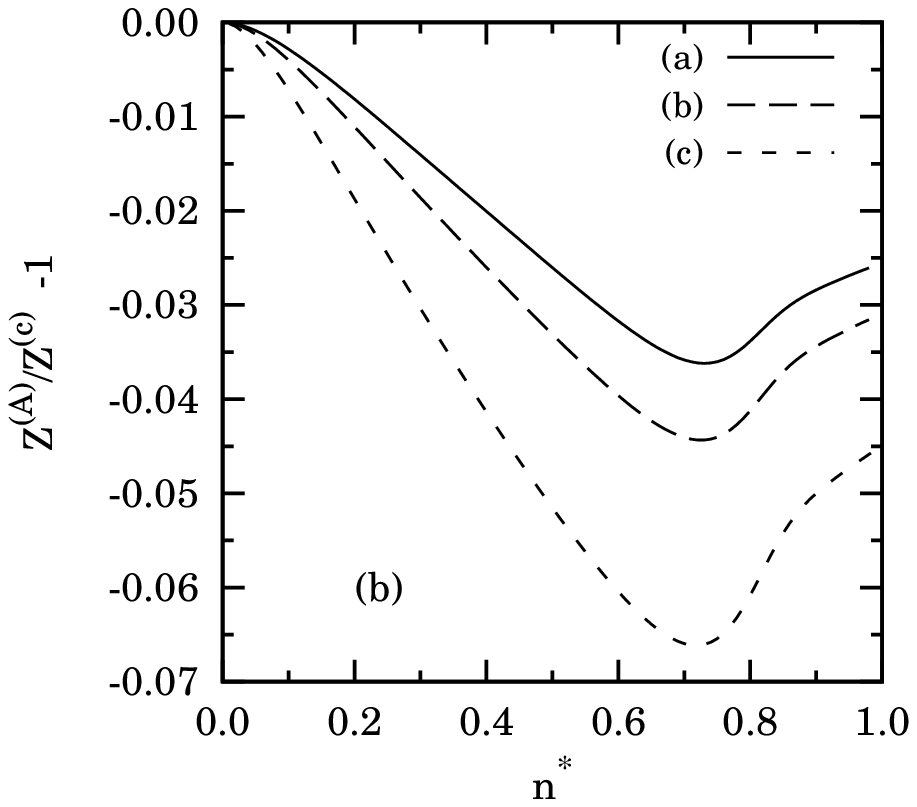}\\
\includegraphics[width=0.45\columnwidth]{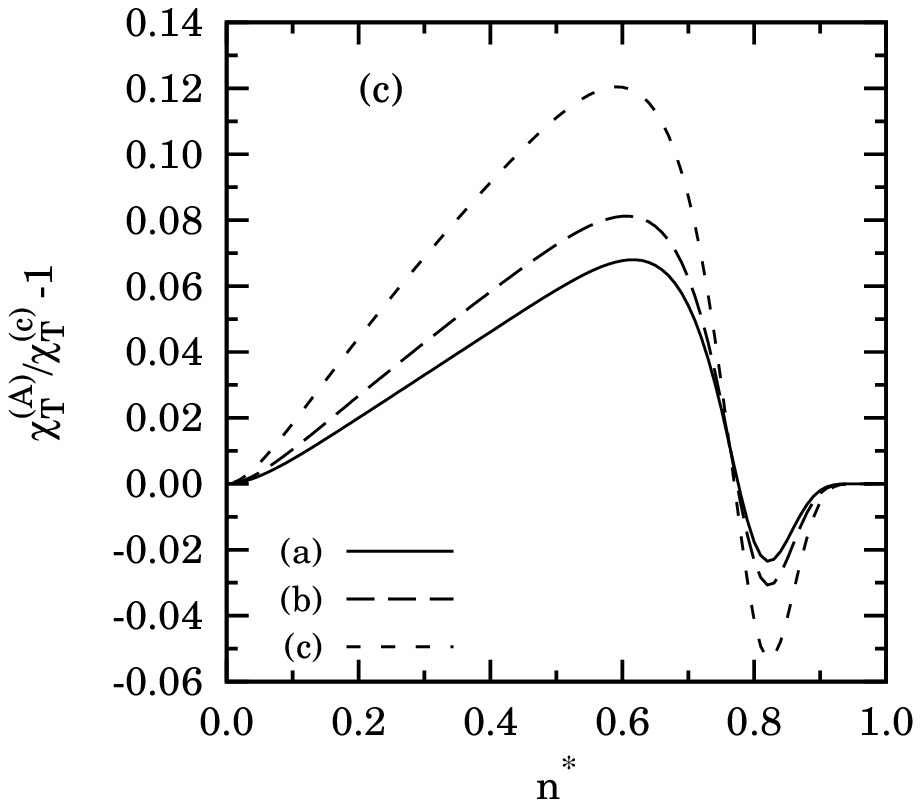}\includegraphics[width=0.45\columnwidth]{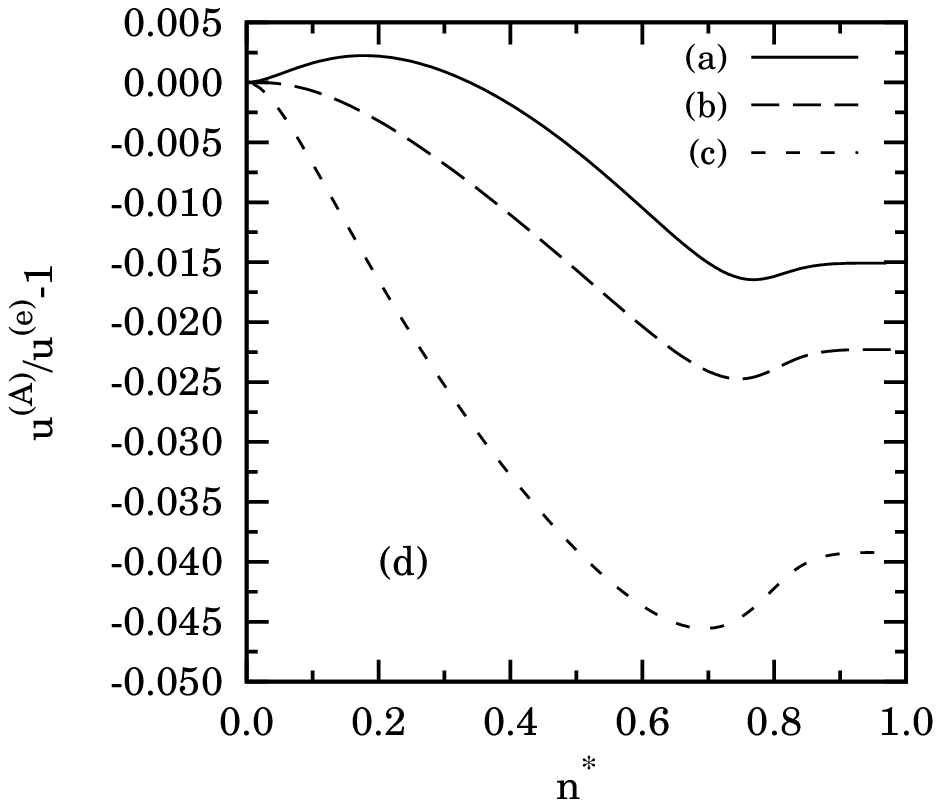}
\end{center}
\caption{Plot of the relative differences (a) $Z^{(\mathrm{A})}/Z^{(\mathrm{v})}-1$, (b) $Z^{(\mathrm{A})}/Z^{(\mathrm{c})}-1$, (c) $\chi_T^{(\mathrm{A})}/\chi_T^{(\mathrm{c})}-1$, and (d) $u^{(\mathrm{A})}/u^{(\mathrm{e})}-1$ as functions of density for the TS fluid ($2\lambda_1=\lambda=3$, $\epsilon_2/\epsilon=\frac{1}{2}$) at $T^*=1$. The curves correspond to the approximations $[1^{(11)}2^{(11)}3^{(11)}]_{a}$ (---), $[1^{(11)}2^{(11)}3^{(11)}]_{b}$ (-- --), and $[1^{(11)}2^{(11)}3^{(11)}]_{c}$ (- - -).}
\label{fig:TSthermod}
\end{figure}

\subsection{Thermodynamic Properties}
As discussed in Sect.\ \ref{sec4.2}, the ``direct'' route $n=\ell/\widetilde{p}_\ell$ allows one to obtain $Z^{(\mathrm{A})}$, $u^{(\mathrm{A})}$, and $\chi_T^{(\mathrm{A})}$. The compressibility route yields $\chi_T^{(\mathrm{c})}$ and $Z^{(\mathrm{c})}$ again from Eqs.\ \eqref{4.7abc} and \eqref{4.8}. As for $u^{(\mathrm{e})}$ and $Z^{(\mathrm{v})}$, the counterparts of Eqs.\ \eqref{4.5} and \eqref{4.6} are
\bq
\label{5.2}
\frac{u^{(\mathrm{e})}}{\epsilon}=-\int_1^{\lambda_1} \rmd r\, \left[p_1(r)+p_2(r)\right]
-\frac{\epsilon_2}{\epsilon}\int_{\lambda_1}^\lambda \rmd r\, \left[p_1(r)+p_2(r)\right],
\eq
\bq
\label{5.3}
Z^{(\mathrm{v})}&=&1+n\left[g(1^+)-\lambda_1\left(\rme^{\beta(\epsilon-\epsilon_2)}-1\right)g(\lambda_1^+)
-\lambda\left(\rme^{\beta\epsilon_2}-1\right)g(\lambda^+)\right]\nn
&=&1+p_1(1^+)-\lambda_1\left(\rme^{\beta(\epsilon-\epsilon_2)}-1\right)\left[p_1(\lambda_1^+)+p_2(\lambda_1^+)\right]
-\lambda\left(\rme^{\beta\epsilon_2}-1\right)\left[p_1(\lambda^+)+p_2(\lambda^+)\right].
\eq
Of course, since $p_2(r)=0$ for $r<2$, the term $p_2(r)$ in the first integral of Eq.\ \eqref{5.2} and the term $p_2(\lambda_1^+)$ in Eq.\ \eqref{5.3} can be removed if $\lambda_1<2$, as happens in our specific case ($\lambda_1=1.5$, $\lambda=3$).

\begin{table}
\caption{Some thermodynamic quantities for the TS fluid ($2\lambda_1=\lambda=3$, $\epsilon_2/\epsilon=\frac{1}{2}$) at several values of $T^*$ and  $n^*$.}
\label{tab:tqTS}
\begin{center}
\begin{tabular}{llllllll}
\noalign{\smallskip}\hline
Method & $Z^{(\mathrm{A})}$ & $Z^{(\mathrm{v})}$ & $Z^{(\mathrm{c})}$& $\chi_T^{(\mathrm{A})}$ & $\chi_T^{(\mathrm{c})}$ &$u^{(\mathrm{A})}/\epsilon$&
$u^{(\mathrm{e})}/\epsilon$\\
\noalign{\smallskip}\hline\noalign{\smallskip}
\multicolumn{8}{c}{$T^*=1, n^*=0.1$}\\
\noalign{\smallskip}\hline\noalign{\smallskip}
MC& \multicolumn{3}{c}{$ 0.9425(5)$} & \multicolumn{2}{c}{---}  &\multicolumn{2}{c}{$-0.2341(6)$}\\
$[1^{(11)}2^{(11)}3^{(11)}]_a$&$0.9440$&$0.9421$&$0.9467$&$1.0964$&$1.0882$&$-0.2324$&$-0.2320$\\
$[1^{(11)}2^{(11)}3^{(11)}]_b$&$0.9443$&$0.9421$&$0.9481$&$1.0954$&$1.0841$&$-0.2319$&$-0.2321$\\
$[1^{(11)}2^{(11)}3^{(11)}]_c$&$0.9445$&$0.9420$&$0.9518$&$1.0927$&$1.0730$&$-0.2306$&$-0.2322$\\
\noalign{\smallskip}\hline\noalign{\smallskip}
\multicolumn{8}{c}{$T^*=1, n^*=0.4$}\\
\noalign{\smallskip}\hline\noalign{\smallskip}
MC& \multicolumn{3}{c}{$1.0075(8)$} & \multicolumn{2}{c}{---}  &\multicolumn{2}{c}{$-0.750(1)$}\\
$[1^{(11)}2^{(11)}3^{(11)}]_a$&$1.0193$&$1.0019$&$1.0401$&$0.7279$&$0.6958$&$-0.7391$&$-0.7404$\\
$[1^{(11)}2^{(11)}3^{(11)}]_b$&$1.0228$&$1.0028$&$1.0501$&$0.7231$&$0.6832$&$-0.7338$&$-0.7420$\\
$[1^{(11)}2^{(11)}3^{(11)}]_c$&$1.0317$&$1.0051$&$1.0761$&$0.7111$&$0.6515$&$-0.7214$&$-0.7459$\\
\noalign{\smallskip}\hline\noalign{\smallskip}
\multicolumn{8}{c}{$T^*=1, n^*=0.7$}\\
\noalign{\smallskip}\hline\noalign{\smallskip}
MC&\multicolumn{3}{c}{$1.905(3)$} & \multicolumn{2}{c}{---}&\multicolumn{2}{c}{$-1.2302(8)$}\\
$[1^{(11)}2^{(11)}3^{(11)}]_a$&$1.9182$&$1.9304$&$1.9895$&$0.1269$&$0.1204$&$-1.2105$&$-1.2290$\\
$[1^{(11)}2^{(11)}3^{(11)}]_b$&$1.9240$&$1.9349$&$2.0126$&$0.1272$&$0.1197$&$-1.2003$&$-1.2300$\\
$[1^{(11)}2^{(11)}3^{(11)}]_c$&$1.9392$&$1.9468$&$2.0761$&$0.1280$&$0.1177$&$-1.1763$&$-1.2325$\\
\noalign{\smallskip}\hline\noalign{\smallskip}
\multicolumn{8}{c}{$T^*=5, n^*=0.1$}\\
\noalign{\smallskip}\hline\noalign{\smallskip}
MC& \multicolumn{3}{c}{$1.0831(1)$} & \multicolumn{2}{c}{---}  &\multicolumn{2}{c}{$-0.1441(2)$}\\
$[1^{(11)}2^{(11)}3^{(11)}]_a$&$1.0832$&$1.0832$&$1.0833$&$0.8493$&$0.8489$&$-0.1445$&$-0.1445$\\
$[1^{(11)}2^{(11)}3^{(11)}]_b$&$1.0832$&$1.0832$&$1.0834$&$0.8493$&$0.8487$&$-0.1445$&$-0.1445$\\
$[1^{(11)}2^{(11)}3^{(11)}]_c$&$1.0832$&$1.0832$&$1.0837$&$0.8493$&$0.8481$&$-0.1444$&$-0.1445$\\
\noalign{\smallskip}\hline\noalign{\smallskip}
\multicolumn{8}{c}{$T^*=5, n^*=0.4$}\\
\noalign{\smallskip}\hline\noalign{\smallskip}
MC& \multicolumn{3}{c}{$1.5287(4)$} & \multicolumn{2}{c}{---}  &\multicolumn{2}{c}{$-0.6090(5)$}\\
$[1^{(11)}2^{(11)}3^{(11)}]_a$&$1.5311$&$1.5304$&$1.5347$&$0.4064$&$0.4038$&$-0.6076$&$-0.6083$\\
$[1^{(11)}2^{(11)}3^{(11)}]_b$&$1.5312$&$1.5304$&$1.5364$&$0.4063$&$ 0.4025$&$-0.6072$&$-0.6084$\\
$[1^{(11)}2^{(11)}3^{(11)}]_c$&$1.5316$&$1.5306$&$1.5415$&$0.4061$&$0.3989$&$-0.6057$&$-0.6085$\\
\noalign{\smallskip}\hline\noalign{\smallskip}
\multicolumn{8}{c}{$T^*=5, n^*=0.7$}\\
\noalign{\smallskip}\hline\noalign{\smallskip}
MC& \multicolumn{3}{c}{$3.003(2)$} & \multicolumn{2}{c}{---} & \multicolumn{2}{c}{$-1.188(1)$}\\
$[1^{(11)}2^{(11)}3^{(11)}]_a$&$3.0270$&$3.0275$&$3.0421$&$0.09520$&$0.09470$&$-1.1845$&$-1.1882$\\
$[1^{(11)}2^{(11)}3^{(11)}]_b$&$3.0272$&$3.0277$&$3.0487$&$0.09522$&$0.09453$&$-1.1826$&$-1.1882$\\
$[1^{(11)}2^{(11)}3^{(11)}]_c$&$3.0279$&$3.0282$&$3.0680$&$0.09526$&$0.09403$&$-1.1771$&$-1.1883$\\
\noalign{\smallskip}\hline
\end{tabular}
\end{center}
\end{table}

\begin{figure}[htbp]
\begin{center}
\includegraphics[width=0.45\columnwidth]{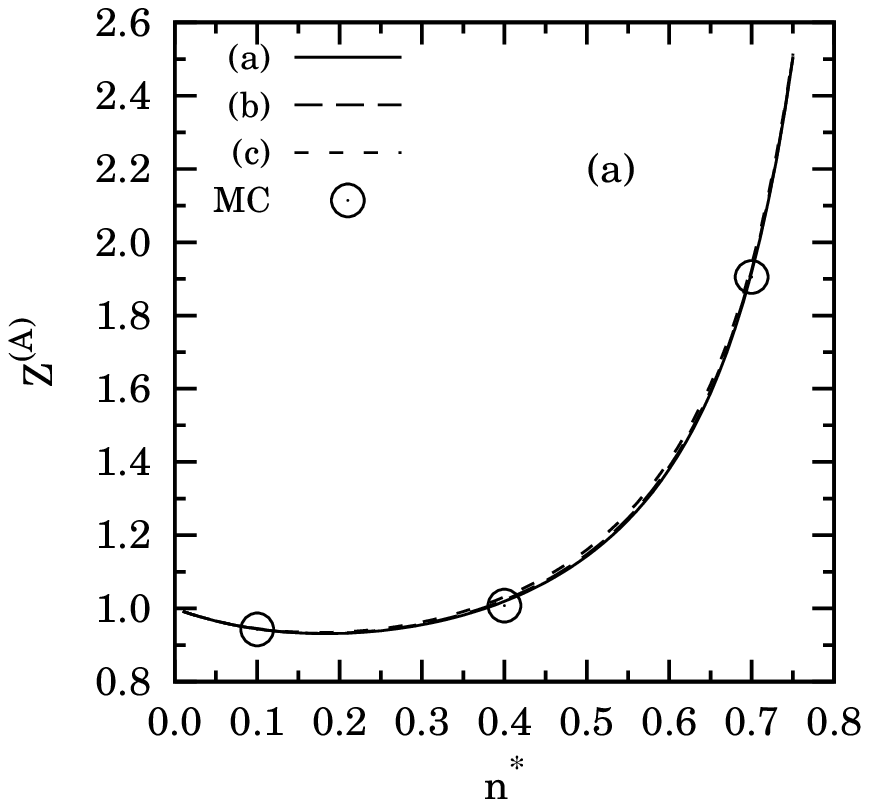}
\includegraphics[width=0.45\columnwidth]{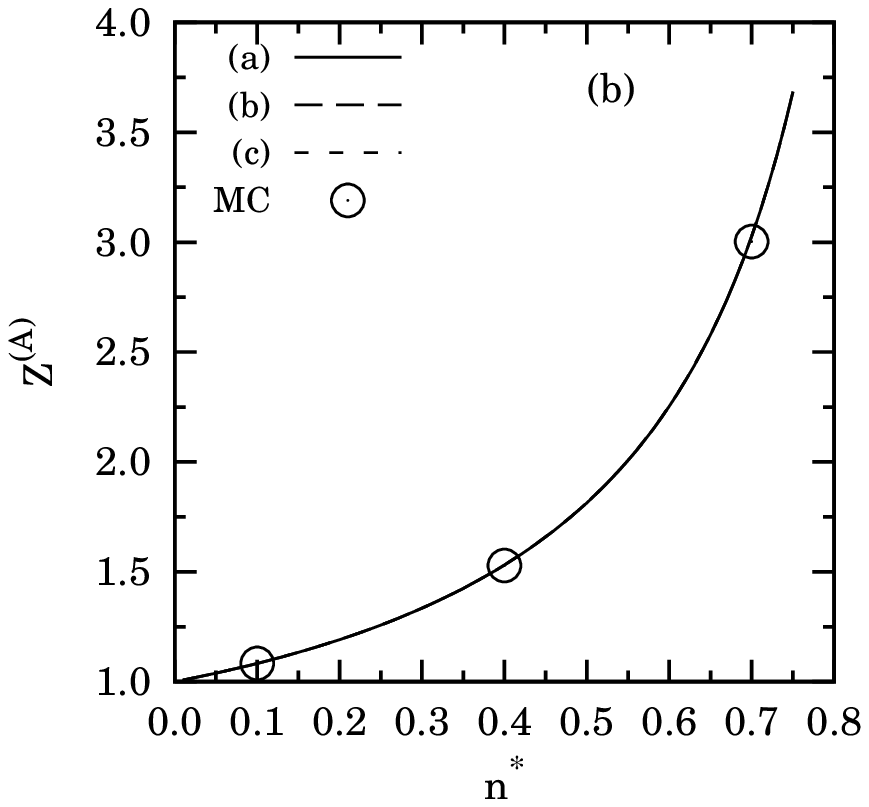}\\
\includegraphics[width=0.45\columnwidth]{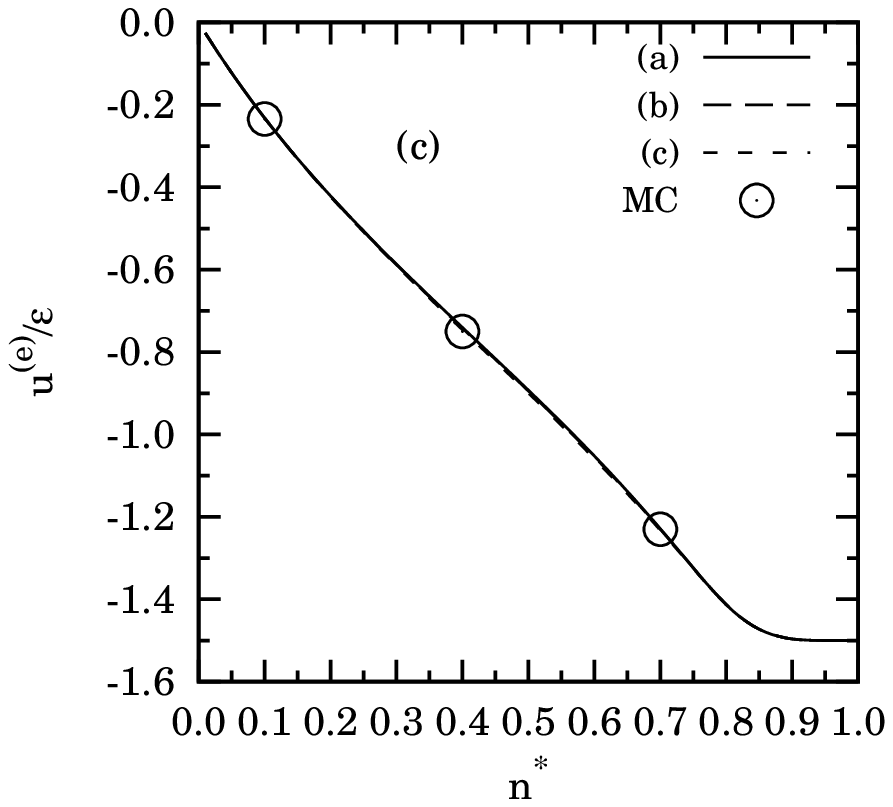}
\includegraphics[width=0.45\columnwidth]{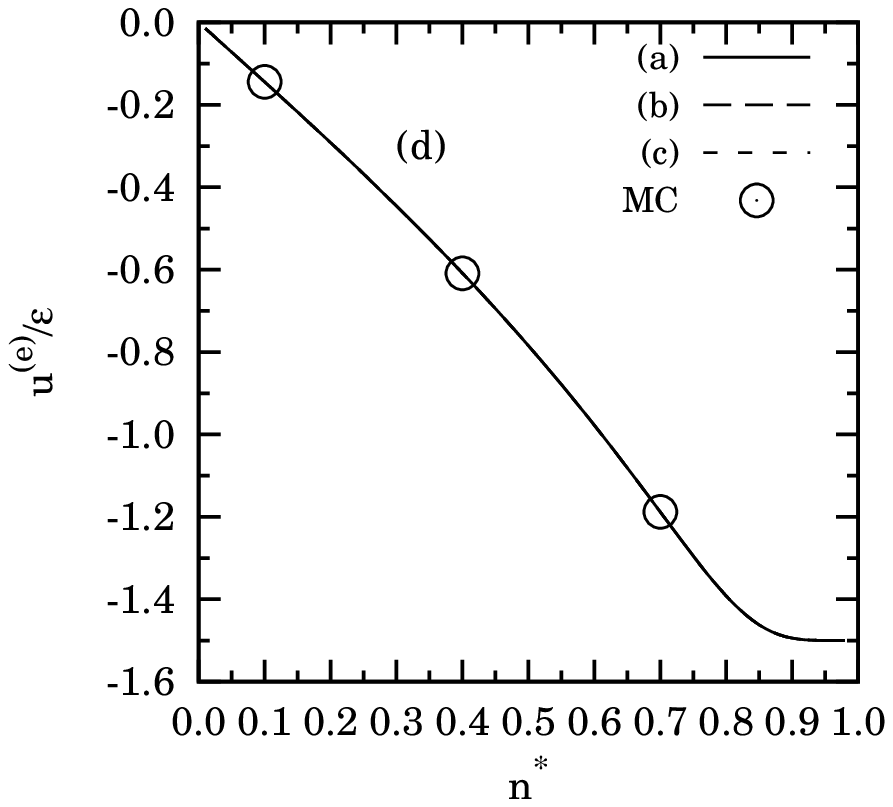}
\end{center}
\caption{Plot of the compressibility factor $Z$ [panels (a) and (b)] and of the
excess internal energy per particle $u$ [panels (c) and (d)] as functions of
density for the TS fluid ($2\lambda_1=\lambda=3$, $\epsilon_2/\epsilon=\frac{1}{2}$) at $T^*=1$ [panels (a) and (c)] and $T^*=5$ [panels (b) and (d)]. The curves are theoretical results $Z^{(\mathrm{A})}$ and $u^{(\mathrm{e})}$, as obtained from the approximations $[1^{(11)}2^{(11)}3^{(11)}]_{a}$ (---), $[1^{(11)}2^{(11)}3^{(11)}]_{b}$ (-- --), and $[1^{(11)}2^{(11)}3^{(11)}]_{c}$ (- - -), while the circles represent MC data.  Note that  the three approximations $[1^{(11)}2^{(11)}3^{(11)}]_{a,b,c}$ yield practically indistinguishable results}
\label{fig:TSthermod1}
\end{figure}

From Fig.\ \ref{fig:TSthermod}  we see again that the approximation $[1^{(11)}2^{(11)}3^{(11)}]_{a}$ is thermodynamically more consistent than $[1^{(11)}2^{(11)}3^{(11)}]_{b}$, and the latter is more consistent than $[1^{(11)}2^{(11)}3^{(11)}]_{c}$. Also comparison between Figs.\ \ref{fig:SWthermod} and \ref{fig:TSthermod} shows that, as expected, our approximations are much more consistent for the TS fluid  than for the SW fluid at common values of $T^*$ and $n^*$.

Table \ref{tab:tqTS} and Fig.\ \ref{fig:TSthermod1} show a comparison with our MC simulation data. The conclusions are similar to those drawn from Table \ref{tab:tq} and Fig.\ \ref{fig:SWthermod1} in the SW case, except that now the performance of the approximations are even better. In fact, from Fig.\ \ref{fig:TSthermod1} one can notice that the three approximations $[1^{(11)}2^{(11)}3^{(11)}]_{a,b,c}$ are practically indistinguishable, even at $T^*=1$.

\section{The Fisher--Widom Line of the Square-Well Model}
\label{sec:WFl}

Rather general arguments \cite{FW69,EHHPS93} suggest a behavior of the one-dimensional $g(r)$ at large
$r$ of the following form,
\begin{eqnarray}
\label{sec:fw1}
g\left(r\right)-1= \sum_{i} A_{i} \rme^{s_i r}
\simeq A_1 \rme^{s_1 r}+A_2 \rme^{s_2 r}+A_3 \rme^{s_3 r}+\cdots,
\end{eqnarray}
where the sum runs over the discrete set of nonzero poles $s_i$ of the Laplace transform $\widehat{G}(s)$, the amplitudes $A_i=\mathrm{Res}\left[\widehat{G}(s)\right]_{s_i}$ being the associated (in general complex) residues, and the ordering $0>\mathrm{Re}(s_1)\geq \mathrm{Re}(s_2)\geq \mathrm{Re}(s_3)\geq \cdots$ is adopted. Note that in Eq.\ \eqref{sec:fw1} the poles are assumed to be single. In case of $s_i$ being a multiple pole, the corresponding term $A_{i} \rme^{s_i r}$ must be replaced by $\mathrm{Res}\left[\rme^{sr}\widehat{G}(s)\right]_{s_i}$. For the discussion below it is sufficient to assume that the pole with the largest real part is single.

Equation \eqref{sec:fw1} shows that the asymptotic decay  of the total correlation function $h(r)=g(r)-1$
is determined by the nature of the pole(s)  with the largest real part of the Laplace transform
$\widehat{G}(s)$ of the RDF. If $s_1=-\kappa+\rmi \omega$ and $s_2=-\kappa-\rmi \omega$ make a pair of complex conjugates, then the asymptotic decay of $h(r)$ is \emph{oscillatory}:
\bq
\label{6.2}
h(r)\approx 2|A_1|\rme^{- \kappa r}\cos(\omega r+\delta),
\eq
where $\delta$ is the argument of $A_{1}$, i.e., $A_{1,2}=|A_1|\rme^{\pm \rmi \delta}$.
On the other hand, if $s_1=-\kappa'$ is a real pole, the decay is \emph{monotonic}, namely
\bq
\label{6.3}
h(r)\approx A_1\rme^{- \kappa' r}.
\eq

In general, the oscillatory decay \eqref{6.2} reflects the correlating effects of the repulsive part of the interaction potential, while the correlating effects of the attractive part are reflected by the monotonic decay \eqref{6.3}. At a given temperature, the first type of decay takes place at sufficiently high values of  pressure (or density), whereas the monotonic decay occurs at sufficiently low values of  pressure (or density). Following Fisher and Widom \cite{FW69}, the locus of transition points from one type to the other one ($\kappa'=\kappa$) defines a line (the so-called FW line) in the pressure (or density) versus temperature plane , with a maximum defining a sort of pseudocritical point.

If the interactions are restricted to the 1st nn, the exact solution is given by Eq.\ \eqref{Gs1}, so that the poles of $\widehat{G}(s)$ are the roots of $\widehat{p}_1(s)-1$ \cite{FW69,FGMS09}. Due to the property \eqref{convs}, those are also roots of $\widehat{p}_\ell(s)-1$ with $\ell=2,3,\ldots$. On the other hand, this equivalence is broken if the interactions involve 2nd nn and we use our approximations \eqref{RDF2ab} and \eqref{RDF3abc}.
Thus, the poles of $\widehat{G}(s)$ are determined by the roots of $\widehat{p}_1(s)-1$ in the approximations \eqref{RDF2b} and \eqref{RDF3c}, the roots of $\widehat{p}_2(s)-1$ in the approximations \eqref{RDF2} and \eqref{RDF3b}, and  the roots of $\widehat{p}_3(s)-1$ in the approximation \eqref{RDF3a}. In each case, the FW line is obtained by solving the set of coupled equations
\begin{subequations}
\label{6.4abc}
\beq
\widehat{p}_\ell(s=-\kappa)-1=0,
\label{6.4a}
\eeq
\beq
\mathrm{Re}\left[\widehat{p}_\ell(s=-\kappa\pm\rmi\omega)\right]-1=0,
\label{6.4b}
\eeq
\beq
\mathrm{Im}\left[\widehat{p}_\ell(s=-\kappa\pm\rmi\omega)\right]=0
\label{6.4c},
\eeq
\end{subequations}
where in Eq.\ \eqref{6.4a} we have taken into account that $\kappa'=\kappa$ on the FW line. At given $T$, the solution to Eqs.\ \eqref{6.4abc} (with $\ell=1$, $2$, or $3$) gives $p$, $\kappa$, and $\omega$ on the FW line.
We see that, as happened with the thermodynamic quantities,
$[1^{(\alpha_1)}2^{(\alpha_2)}]_a=[1^{(\alpha_1)}2^{(\alpha_2)}3^{(\alpha_2)}]_b$ and $[1^{(\alpha_1)}2^{(\alpha_2)}]_b=[1^{(\alpha_1)}2^{(\alpha_2)}3^{(\alpha_3)}]_c$ in what concerns the FW line. Again, we can focus on  the three approximations $[1^{(11)}2^{(11)}3^{(11)}]_{a,b,c}$.

\begin{figure}
\begin{center}
\includegraphics[width=0.45\columnwidth]{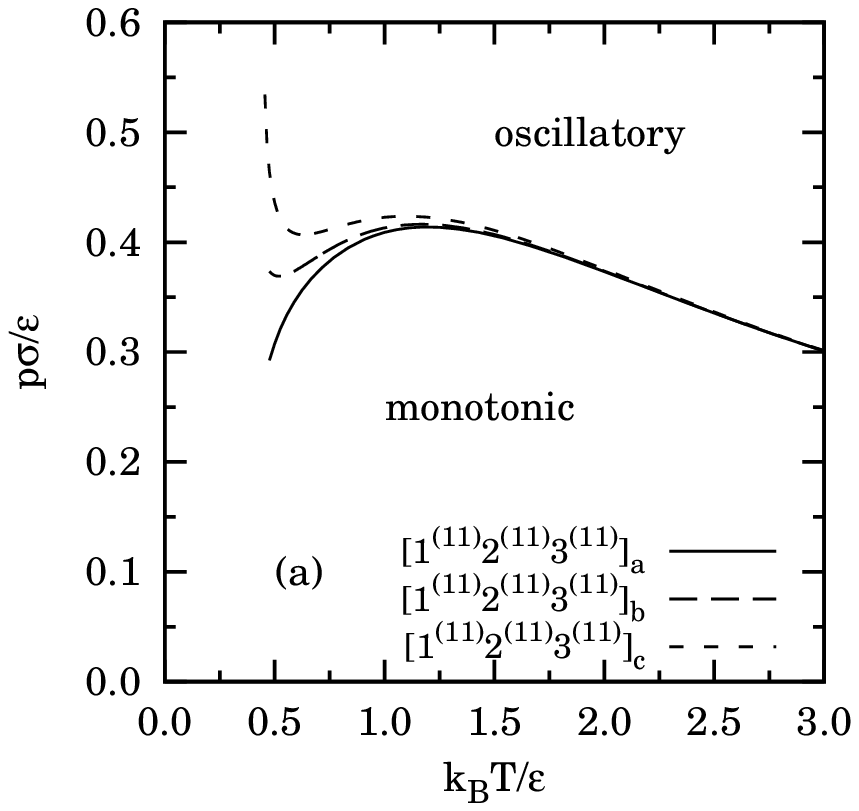}\includegraphics[width=0.45\columnwidth]{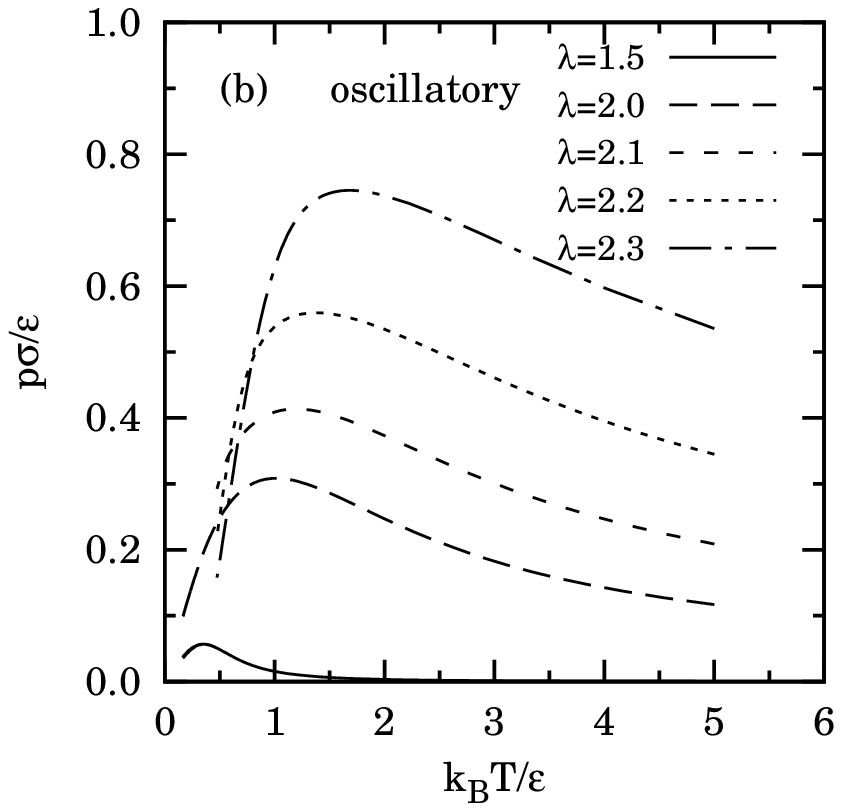}
\end{center}
\caption{(a) FW line for the SW fluid with
  $\lambda=2.1$, as  calculated from the approximations $[1^{(11)}2^{(11)}3^{(11)}]_{a}$ (---), $[1^{(11)}2^{(11)}3^{(11)}]_{b}$ (-- --), and $[1^{(11)}2^{(11)}3^{(11)}]_{c}$ (- - -). (b) FW lines for the SW fluid with
  $\lambda=1.5$ (---), $2$ (-- --), $2.1$ (- - -), $2.2$ ($\cdots$), and $2.3$ (-- $\cdot$ --); the lines corresponding to $\lambda=1.5$ and $2$ are exact, while those corresponding to $\lambda>2$ were calculated from the approximation $[1^{(11)}2^{(11)}3^{(11)}]_{a}$.}
\label{fig:wf311}
\end{figure}

Now we apply the above general description to the SW potential \eqref{SWpotential} with varying range $\lambda$.
In Fig.\ \ref{fig:wf311}(a) we show a comparison between the three approximations $[1^{(11)}2^{(11)}3^{(11)}]_{a,b,c}$ at $\lambda=2.1$.
They agree well up to approximately the location of the pseudocritical point, i.e., for $T^*\gtrsim 1$. At lower temperatures, however, the two approximations $[1^{(11)}2^{(11)}3^{(11)}]_{b,c}$ [i.e., the solutions to Eqs.\ \eqref{6.4abc} with $\ell=1,2$] exhibit an unphysical increase of the pressure as temperature decreases. This may be regarded
as an artifact of the approximations, which break down at small
temperatures when the particles of the fluid are highly coupled. Nevertheless, the approximation $[1^{(11)}2^{(11)}3^{(11)}]_{a}$ is qualitatively correct.

The influence of $\lambda$ on the FW line is analyzed in Fig.\ \ref{fig:wf311}(b), where the exact results for $\lambda=1.5$ and $2$ are contrasted with those resulting from our approximation $[1^{(11)}2^{(11)}3^{(11)}]_{a}$ for $\lambda=2.1$, $2.2$, and $2.3$.
We can see that the most relevant trends observed when increasing $\lambda$ in the ``safe'' exact domain ($\lambda\leq 2$) are extended to the domain $\lambda>2$. In particular, the location of the pseudocritical point moves to larger values of temperature and, especially, of pressure as the range $\lambda$ increases. Also, at a given $T^*$ (larger than the pseudocritical value),  the oscillatory--monotonic transition takes place at noticeably higher values of $p^*$, even if $\lambda$ is increased very little. Nonetheless, the approximation $[1^{(11)}2^{(11)}3^{(11)}]_{a}$ predicts too sharp a decay of the FW line below the pseudocritical temperature, even crossing the lines of smaller $\lambda$. We believe this to be an artifact of the approximation, which becomes less reliable as temperature decreases.

\section{Conclusions}
\label{sec:conclusions}

In this work we have proposed a detailed analysis of approximate analytical extensions
to 2nd nn fluids of the exact analytical solution of 1st nn fluids confined in one spatial dimension.
The inclusion of the 2nd nn interactions renders the calculation of the
partition function and the various correlation functions extremely more
cumbersome than in the 1st nn case. In particular, the exact solution is not tractable anymore. A detailed diagrammatic analysis of the exact structure of the correlation functions and of their various approximations has also been carried out in the
spirit of the Mayer cluster diagrams.

Two stages have been followed to determine the RDF $g(r)$. In the first stage, attention is focused on the $\ell$th nn probability distribution function $p_\ell(r)$. The exact $p_\ell (r)$, which involves a many-body problem, is approximated by $p_\ell^{(k_1k_2)}(r)$, where only integrals involving the $k_1$ particles to the left of particle $1$ and the $k_2$ particles to the right of particle $\ell+1$ are incorporated.
In the second stage, a \emph{finite} number of functions $\widehat{p}_1 (s)$, $\widehat{p}_2 (s)$, \ldots, $\widehat{p}_\ell (s)$ (in Laplace space) are used to approximate the Laplace transform of the RDF, $\widehat{G}(s)$.
This double sequence of approximations  becomes the exact
solution only in the infinite order limit, i.e., if $\ell\to\infty$ in the construction of  $\widehat{G}(s)$ and  $k_1,k_2\to\infty$ in the construction of $p_\ell^{(k_1k_2)}(r)$. Here we have restricted ourselves to $k_1,k_2\leq 1$ in the construction of $p_\ell^{(k_1k_2)}(r)$ and to $\ell=3$ in the construction of $\widehat{G}(s)$. Out of this, our recommended approximation is given by Eq.\ \eqref{RDF3a} complemented by Eqs.\ \eqref{(11)_1}, \eqref{(11)_2}, and \eqref{111}. We have denoted this combined approximation as $[1^{(11)}2^{(11)}3^{(11)}]_{a}$.

Our theoretical approach has been assessed by comparison with our own MC simulations for the SW and TS fluids, in both cases with the largest potential range compatible with 2nd nn interactions. The comparison has been made both at the level of the most common thermodynamic
quantities, such as the equation of state and the internal energy per
particle, and at the level of the RDF.
Also some internal thermodynamic consistency tests (three different routes
to the equation of state, two to the isothermal susceptibility, and two to the internal energy) have been carefully addressed. We have found that
 $[1^{(11)}2^{(11)}3^{(11)}]_{a}$ is a sufficiently good approximation for the
SW fluid, while the simpler approximation $[1^{(11)}2^{(11)}]_{a}$ is already  good enough for the
TS potential (at least when the depth of the second step, affecting also
the 2nd nn, is one half the depth of the first step,
affecting only the 1st nn).

Finally, we have calculated the FW line (separating states where the correlation function decays monotonically from those where it decays in an oscillatory way) of the SW model for various ranges. Although the reliability of the approximation $[1^{(11)}2^{(11)}3^{(11)}]_{a}$ is expected to worsen at temperatures lower than the pseudocritical one, the results clearly show that the FW line is rather sensitive to changes in the potential range, the pseudcritical point moving to higher temperatures and, especially, pressures as the range increases

The analysis presented here can become a useful tool, as an approximate extension to the 2nd nn fluid of
the exact 1st nn fluid analytical solution, whenever one wants to find an easy, albeit
approximate, solution for the fluid properties, both structural and thermodynamic. It can avoid having to
resort to  simulations or serve as a guide to them, with the necessary caution of keeping in
mind that the reliability of the approach is expected to worsen at very low temperatures.

Our general scheme can be easily generalized to the inclusion of
any number of nn interactions, but one must treat each case independently.
This is an alternative procedure
to the eigenvalue route used, for example, in Ref.\ \cite{F16} towards the analytical solution of a generic (non-quantum) one-dimensional fluid of impenetrable particles interacting through pair interactions which  reflect the fact that the particles are ``living'' on the line or simply moving on the line but embedded in a higher dimensional space.
It is also worth mentioning that the isothermal-isobaric ensemble results can be regarded as evaluation of a generating function; embedding in an overcomplete density functional formalism \cite{P02}  makes extension to non-1st nn interactions possible.

On the other hand, the method cannot be easily
generalized to more than one spatial dimension since a crucial
ingredient is the ordering of the particles on the line, which is lost
in dimensions higher than one.
Of course, following a bottom-up strategy, one is free to blindly adapt
our approximation scheme, for example, to the more realistic three-dimensional case (where even the SW potential with $\lambda<2$ cannot be solved
exactly), but the result remains uncertain.

\begin{acknowledgements}
R.F. is grateful to the Departamento de F\'isica, Universidad de Extremadura, for its hospitality during a two-month stay in early 2017, when this work was initiated. A.S. acknowledges the financial support of the
Ministerio de Econom\'ia y Competitividad (Spain) through Grant No.\ FIS2016-76359-P and  the Junta de Extremadura (Spain) through Grant No.\ GR15104, both partially financed by ``Fondo Europeo de Desarrollo Regional'' funds.

\end{acknowledgements}

\appendix

\section{Derivation of Eqs.\ \eqref{p1(r)2nn}, \eqref{p2exact}, and \eqref{p3exact}}
\label{app:A}

\subsection{Equation \eqref{p1(r)2nn}}
By  a change of variables from absolute
coordinates ($\{x_i\}$) to relative coordinates ($\{r_i\}$), we find that Eq.\ \eqref{p1(r)} can be rewritten as
\bq
p_1(r)&\propto& \rme^{-\beta\phi(r)}\int_r^\infty \rmd L\,\rme^{-\beta pL}
\int_0^{L-r}\rmd r_3\,\rme^{-\beta\phi(r_3)}\rme^{-\gamma\beta\phi(r+r_3)}
\int_0^{L-r-r_3}\rmd r_4\,\rme^{-\beta\phi(r_4)}\rme^{-\gamma\beta\phi(r_3+r_4)}\cdots\nn
&&\times\int_0^{L-r-r_3-\cdots-r_{N-1}}\rmd r_N\,\rme^{-\beta\phi(r_N)}
\rme^{-\gamma\beta\phi(r_{N-1}+r_N)} \rme^{-\beta\phi(r_{N+1})}\rme^{-\gamma\beta\phi(r_N+r_{N+1})}\rme^{-\gamma\beta\phi(r_{N+1}+r)},
\eq
where $r_{N+1}=L-r-r_3-\ldots-r_N$ and we have taken into account that $r_{N+2}=r$.
The change of variables $L\to L'=L-r$ implies that a factor
$\rme^{-\beta pr}$ comes out of the integrals. Exchanging the integral
over $L'$ and the integral over $r_3$ we get
\bq
p_1(r)&\propto& \rme^{-\beta [pr+\phi(r)]}
\int_0^\infty \rmd r_3\,\rme^{-\beta\phi(r_3)}\rme^{-\gamma\beta\phi(r+r_3)}
\int_{r_3}^\infty \rmd L'\,\rme^{-\beta pL'}\int_0^{L'-r_3}\rmd r_4\,\rme^{-\beta\phi(r_4)}\rme^{-\gamma\beta\phi(r_3+r_4)}\cdots
\nn
&&\times\int_0^{L'-r_3-\cdots-r_{N-1}}\rmd r_N\,\rme^{-\beta\phi(r_N)}
\rme^{-\gamma\beta\phi(r_{N-1}+r_N)} \rme^{-\beta\phi(r_{N+1})}\rme^{-\gamma\beta\phi(r_N+r_{N+1})}\rme^{-\gamma\beta\phi(r_{N+1}+r)}.
\eq
Next, changing variables $L'\to L''=L'-r_3$ and exchanging the integral
over $L''$ and the integral over $r_4$ we find
\bq
p_1(r)&\propto& \rme^{-\beta [pr+\phi(r)]}
\int_0^\infty \rmd r_3\,\rme^{-\beta[pr_3+\phi(r_3)]}\rme^{-\gamma\beta\phi(r+r_3)}
\int_0^{\infty}\rmd r_4\,\rme^{-\beta\phi(r_4)}\rme^{-\gamma\beta\phi(r_3+r_4)}
\int_{r_4}^\infty \rmd L''\,\rme^{-\beta pL''}\cdots
\nn
&&\times\int_0^{L''-r_4-\cdots-r_{N-1}}\rmd r_N\,\rme^{-\beta\phi(r_N)}
\rme^{-\gamma\beta\phi(r_{N-1}+r_N)} \rme^{-\beta\phi(r_{N+1})}\rme^{-\gamma\beta\phi(r_N+r_{N+1})}\rme^{-\gamma\beta\phi(r_{N+1}+r)}.
\eq
This process can be continued with $L''\to L'''=L''-r_4$, $L'''\to L^{\mathrm{IV}}=L'''-r_5$, \ldots, until arriving to $L^{(N-1)}=L-r-r_3-\cdots-r_N=r_{N+1}$ (see Fig.\ \ref{fig:diag}). After performing all these changes it is easy to see that Eq.\ \eqref{p1(r)2nn} is finally obtained.

\subsection{Equation \eqref{p2exact}}
Using  Eq.\ (\ref{pp}) and the pbc, Eq.\ \eqref{p2(r)} can be rewritten as
\bq
p_2(\rp)&\propto& \rme^{-\gamma\beta\phi(\rp)}\int_{\rp}^\infty \rmd L\,\rme^{-\beta pL}\int_0^{\rp}
\rmd r_2\,\rme^{-\beta\phi(r_2)}\rme^{-\beta\phi(\rp-r_2)}\int_0^{L-\rp}\rmd r_4\,\rme^{-\beta\phi(r_4)}\rme^{-\gamma\beta\phi(\rp-r_2+r_4)}\int_0^{L-\rp-r_4}\rmd r_5\,\rme^{-\beta\phi(r_5)}\rme^{-\gamma\beta\phi(r_4+r_5)}\cdots
\nn
&&\times\int_0^{L-\rp-r_4-\cdots-r_{N-1}}\rmd r_N\,\rme^{-\beta\phi(r_N)}
 \rme^{-\gamma\beta\phi(r_{N-1}+r_N)}\rme^{-\beta\phi(r_{N+1})}\rme^{-\gamma\beta\phi(r_N+r_{N+1})}\rme^{-\gamma\beta\phi(r_{N+1}+r_2)}.
\eq
Analogously to the case of $p_1(r)$, the change of variables $L\to L'=L-\rp$ implies that a factor
$\rme^{-\beta p \rp}$ comes out of the integrals. Exchanging the integral
over $L'$ and the integral over $r_4$ we get
\bq
p_2(\rp)&\propto& \rme^{-\beta[ p \rp+\gamma\phi(\rp)]}\int_0^{\rp}
\rmd r_2\,\rme^{-\beta\phi(r_2)}\rme^{-\beta\phi(\rp-r_2)}\int_0^\infty \rmd r_4\,\rme^{-\beta\phi(r_4)}\rme^{-\gamma\beta\phi(\rp-r_2+r_4)}\int_{r_4}^\infty \rmd L'\,\rme^{-\beta pL'}
\int_0^{L'-r_4}\rmd r_5\,\rme^{-\beta\phi(r_5)}\rme^{-\gamma\beta\phi(r_4+r_5)}\cdots
\nn
&&\times\int_0^{L'-r_4-\cdots-r_{N-1}}\rmd r_N\,\rme^{-\beta\phi(r_N)}
\rme^{-\gamma\beta\phi(r_{N-1}+r_N)} \rme^{-\beta\phi(r_{N+1})}\rme^{-\gamma\beta\phi(r_N+r_{N+1})}\rme^{-\gamma\beta\phi(r_{N+1}+r_2)}.
\eq
Successive changes of  variables $L'\to L''=L'-r_4$, $L''\to L'''=L''-r_5$, $L'''\to L^{\mathrm{IV}}= L'''-r_6$, \ldots, until $L^{(N-2)}=L-\rp-r_4-\cdots-r_N=r_{N+1}$ allows one to derive Eq.\ \eqref{p2exact}.

\subsection{Equation \eqref{p3exact}}
As before, use of  Eq.\ (\ref{pp}) and of the pbc yields
\bq
p_3(\rpp)&\propto&\int_{\rpp}^\infty \rmd L\,\rme^{-\beta pL}\int_0^{\rpp}
\rmd r_2\,\rme^{-\beta\phi(r_2)}\rme^{-\gamma\beta\phi(\rpp-r_2)}\int_0^{\rpp-r_2}\rmd r_3\,
\rme^{-\beta\phi(r_3)}\rme^{-\gamma\beta\phi(r_2+r_3)}\rme^{-\beta\phi(\rpp-r_2-r_3)}\int_0^{L-\rpp}\rmd r_5\,\rme^{-\beta\phi(r_5)}\rme^{-\gamma\beta\phi(\rpp-r_2-r_3+r_5)}
\nn
&&\times\int_0^{L-\rpp-r_5}\rmd r_6\,\rme^{-\beta\phi(r_6)}\rme^{-\gamma\beta\phi(r_5+r_6)}\cdots \int_0^{L-\rpp-r_5-\cdots-r_{N-1}}\rmd r_N\,\rme^{-\beta\phi(r_N)}
\rme^{-\gamma\beta\phi(r_{N-1}+r_N)}
\rme^{-\beta\phi(r_{N+1})}\rme^{-\gamma\beta\phi(r_N+r_{N+1})}\rme^{-\gamma\beta\phi(r_{N+1}+r_2)}.\nn
\eq
Again, the change of variables $L\to L'=L-\rpp$ implies that a factor
$\rme^{-\beta p \rpp}$ comes out of the integrals. Exchanging the integral
over $L'$ and the integral over $r_5$, changing variables to $L'\to
L''=L'-r_5$, and continuing this process we finally reach Eq.\ \eqref{p3exact}.






\end{document}